\documentclass[prd,nofootinbib,preprint]{revtex4-2}
\usepackage[utf8]{inputenc}
\usepackage{amsmath}
\usepackage{amssymb}
\usepackage{textcomp}
\usepackage{array,multirow}
\usepackage{slashed}
\usepackage{tabularx}
\usepackage{hyperref}
\usepackage{mathtools}
\usepackage{natbib}
\usepackage{graphicx}
\usepackage{xcolor}
\usepackage{textcomp}
\usepackage{mathrsfs}
\usepackage[labelformat=parens]{subcaption}
\usepackage{varwidth}
\newcommand{\amp}{\mathcal{A}}
\newcommand{\thalf}{t_{1/2}^{0\nu\beta\beta}}
\begin{document}
\title{Dark matter motivated sterile neutrino contribution to neutrinoless 
double beta decay}
\author{Debashree Priyadarsini Das}
\email{debashreepriyadarsini\_das@nitrkl.ac.in}
\author{Sasmita Mishra}%
\email{mishras@nitrkl.ac.in}
\affiliation{%
Department of Physics and Astronomy, National Institute of Technology Rourkela, Odisha, India, 769008}%
\begin{abstract}
The exact seesaw relation in a type-I seesaw framework puts constraints on the relations between
active and sterile neutrino sectors in terms of their masses and mixing angles. In such a setup, we employ a model-independent approach to investigate the signature of sterile neutrinos in the half-life of the neutrinoless double beta ($0\nu\beta\beta$) decay process. In particular, we aim to study the contribution of sterile neutrinos
in the mass range $\sim$~keV that is motivated by the dark matter constituent of the Universe. 
Further, the masses of the sterile neutrinos are determined by the active neutrino masses, mixing angles, and phases, and active-sterile mixing angles and $CP$-violating phases. The parameter space is constrained by the
exact seesaw relation, thereby making the analysis constrained. After capturing the parameter space that can account for $\sim~$keV scale masses for the sterile neutrinos, we adopt the chiral effective field theory approach to calculate the half-life and effective mass in the $0\nu\beta\beta$ decay. 
As the study transitions from the TeV scale to scenarios involving at least one sterile neutrino in the keV mass range, it reveals a significant modification of the effective mass. In particular, the cancellation region associated with the normal mass hierarchy for TeV-scale sterile neutrinos no longer persists when a keV-scale sterile neutrino is introduced, resulting in a finite effective mass that future experiments can probe. Likewise, the involvement of keV-scale sterile neutrino in the inverted mass hierarchy case makes the band distorted and scattered points appear around the main band.
\end{abstract}
\maketitle
\newpage
\section{Introduction}\label{intro}
The extension of the Standard Model (SM) through singlet right-handed (RH) neutrinos is recognized as an attractive approach where the RH neutrinos not only generate masses for SM neutrinos through the type-I seesaw mechanism, but also enable their Majorana nature \cite{Minkowski:1977sc, Mohapatra:1979ia, Yanagida:1980xy}. 
Their masses remain unconstrained by the electroweak symmetry breaking, and can span from $\sim$ eV to several TeV or even match the GUT scale. RH neutrinos, usually referred to as sterile neutrinos with mass $\gg {\rm eV}$ are often addressed as Heavy Neutral Leptons (HNL) \cite{Abdullahi:2022jlv}. Sterile neutrinos in the mass range between 10 GeV and the electroweak scale are being targeted for their direct search in the collider experiments at LHC \cite{ATLAS:2019kpx, DeVries:2020jbs} and LEP \cite{Antusch:2016ejd, Yang:2023ice}. It is noteworthy that the addition of sterile neutrinos is found to be advantageous from various perspectives, including leptogenesis and the composition of dark matter. They can account for the anomalies observed from several oscillation experiments in their eV-scale mass range \cite{LSND:2001aii, MiniBooNE:2007uho, Mention:2011rk, Abdurashitov:1996dp}. Sterile neutrinos with masses in the keV range are an ideal candidate for warm dark matter as identified in Ref.\cite{Dodelson:1993je}. 
The $3.5$ keV line in the X-ray spectra from various astronomical observations 
\cite{Iakubovskyi:2015dna,Boyarsky:2014jta,Boyarsky:2014ska,Ruchayskiy:2015onc,Bulbul:2014sua} can 
be interpreted as the decay of DM particle with mass $\sim 7$ keV, particularly a sterile neutrino with the mixing angle $\sin^2 \theta \simeq (0.2 - 2) \times 10^{-10}$. Sterile neutrinos in the keV mass range can be produced non-thermally in the early universe due to their feeble interaction. In the non-thermal resonant production scenario (Dodelson-Widrow mechanism) \cite{Merle:2015vzu, Dodelson:1993je}, they can be produced via oscillations of active-sterile neutrinos in the primordial plasma at temperatures of around a few hundred MeV. As the production rate depends on the active-sterile mixing, so with a decrease in the Universe's temperature, the active-sterile mixing angle increases and consequently their oscillation become more efficient. On the other hand, in the case of a resonant production scenario, the presence of a large lepton asymmetry triggers MSW like resonance that increases the active-sterile oscillation efficiency even at small mixing angles \cite{Shi:1998km}. 
In this work, we investigate the role of such sterile neutrinos in neutrinoless double beta decay ($0\nu\beta \beta$) within an effective-field-theory framework.

Sterile neutrinos, being Majorana in nature, inherently violate lepton number through their self-conjugate mass terms, and hence can offer a natural source for lepton number-violating (LNV) processes. In the context of LNV processes, $0\nu\beta \beta$ comes as a priori of interest due to its ability to ascertain the Majorana nature of neutrinos and to untangle the absolute neutrino mass scale \cite{Avignone:2007fu, Giuliani:2012zu}. Due to advancements in their experimental probe, the parameter space, including neutrino masses, mixing angles, and $ CP$-violating phases, is receiving precise constraints \cite{Dolinski:2019nrj, DellOro:2016tmg}. The direct observation of this decay will not only provide compelling evidence for lepton number non-conservation directly, but also lend indirect evidence to leptogenesis through its parameter space \cite{Harz:2021psp, deVries:2024rfh}. The quantity that is being measured through the $0\nu \beta \beta$ experiments is the inverse of the half-life period of the decaying isotopes, i.e., $\left(\thalf\right)^{-1}$. Among the present generation experiments, the most stringent bounds on $\thalf$ are obtained from GERDA ($> 1.8 \times 10^{26}$ year) \cite{GERDA:2020xhi} and KamLAND-Zen ($> 2.3 \times 10^{26}$ year) \cite{KamLAND-Zen:2022tow}. An improvement of about two orders of magnitude is expected from various next-generation experiments like nEXO \cite{nEXO:2021ujk}
LEGEND \cite{LEGEND:2021bnm} and CUPID \cite{CUPID:2022wpt}.

Some related work in this direction can be found in Ref. \cite{Bezrukov:2005mx}, where the dark matter motivated sterile neutrinos were found to give a vanishing contribution to effective neutrino mass ($m_{ee}$) in $\nu$MSM \cite{Asaka:2005an}. Also, it was shown in Ref. \cite{Blennow:2010th} that the extra states can also have a significant impact by cancelling the SM neutrino contribution for masses lighter than the nuclear scale and leading to vanishing $ 0\nu\beta\beta$ decay amplitudes even if neutrinos are Majorana particles.  The calculation was improved in Ref. \cite{Merle:2013ibc} by considering that the sterile neutrinos constitute the full and partial dark matter density, which can give a non-vanishing contribution to the $0\nu\beta\beta$ decay. In the presence of sterile neutrinos in the
mass range ${\mathcal{O}}(1)$ eV to $5$ GeV, the rates for $ 0\nu\beta\beta$ processes would be enhanced due to their resonant contribution, as shown in various works \cite{Atre:2009rg, Cabrera:2023rcy, Bolton:2019pcu}. In this work, taking into consideration the appropriate scale for a systematic chiral effective field theory ($\chi$EFT) formulation, we propose to concentrate on the low-scale seesaw regime. In the canonical high-scale seesaw limit, sterile neutrinos will be integrated out completely, whereas the low-scale seesaw limit carries out a consistent $\chi$EFT treatment by retaining the sterile neutrino propagators explicitly. Additionally, the low-scale seesaw provides distinct phenomenological advantages in terms of its experimental accessibility \cite{Agrawal:2021dbo,Drewes:2013gca,
Kersten:2007vk,Shaposhnikov:2006nn}. It can explain the tinyness of SM neutrinos in a testable framework while keeping the added sterile neutrino masses within a few TeV. In this mass regime, sterile neutrinos can take an active role in the $0\nu \beta \beta$ decay and can account for quantities like its half-life $\thalf$ and $m_{ee}$. This will furnish observable signatures of the decay and valuable insight into the LNV interactions and laboratory experiments like charged lepton flavor violation. Essentially, the experimentally testable link between neutrino physics, cosmology, and collider phenomenology can be established through the sterile neutrinos in the low-scale seesaw limit.

In this work, we add three sterile neutrinos corresponding one-to-one with the three active neutrinos. 
They can be included in the neutrino mixing sector through a $6\times6$ unitary mixing matrix following the procedure considered in the references \cite{Das:2023aic, Xing:2007zj, Xing:2011ur}. By doing this, we can get an extended parameter space spanned by the active-sterile mixing angles, $ CP$-violating phases, and the active neutrino masses, which will act as input variables for the analytical expressions for the masses of the added sterile neutrinos. The parameter space is highly constrained by the exact seesaw relation in the type-I seesaw
framework. We then use these analytical expressions of the masses of the sterile neutrinos and calculate the $0\nu \beta \beta$ decay half-life and effective neutrino mass in the $\chi$EFT framework. The $\chi$EFT approach allows a thorough exploration of the contribution from the NMEs. Following the references \cite{deVries:2024rfh, Dekens:2024hlz}, we have adapted an effective formula for the amplitude, incorporating a summary of all the contributions from both active neutrinos and sterile neutrinos from different momentum regions as per the $\chi$EFT approach. The decay amplitudes, defined according to the $\chi$EFT for different momentum regions, are actually expressed as a coherent combination of various NMEs encapsulating the underlying nuclear structure effects. These NMEs are obtained through many-body nuclear calculations based on various nuclear models. In our work, we consider the values computed from the nuclear shell model for  ${}^{136}\rm{Xe}$ isotope, which are then used in the amplitude expressions to evaluate the $\thalf$ and $m_{ee}$ for the isotope.

The rest of the paper is organized as follows. In section (\ref{sec2}), we present the \texorpdfstring{$3+3$}{} model in the context of type-I seesaw. The incorporation of three sterile states in the lepton sector through the $6\times6$ mixing matrix and the analytical expressions for their mass state in the limit of the exact seesaw relation are described in this section. Section (\ref{sec:ovbb-eft}) is devoted to the analysis half-life $\left(\thalf\right)$ and effective neutrino mass $(m_{ee})$ of $0\nu\beta\beta$ decay within the framework of $\chi$EFT. This section also highlights the effective formula for $0\nu\beta\beta$ decay amplitude $\amp$ as a function of the mass of the mediating neutrino. The parameter scan and necessary numerical results are presented in section (\ref{result-eft}) followed by the conclusion in section (\ref{4-con}). 

\section{Type-I seesaw and the \texorpdfstring{$3+3$}{} model}\label{sec2}
Considering $N$ number of sterile neutrinos as the SM extension, the relevant part of the Lagrangian is given by
\begin{equation}
-{\mathcal{L}} \supset i \overline{N_R}_{i}\gamma^\mu\partial_\mu {N_R}_{i} + \left( Y_{\alpha i} \overline{L}_\alpha \varPhi {N_R}_{i} 
+ \frac{{M_R}_i}{2} \overline{{N^c_R}_{i}}{N_R}_{i} + {\rm h.c.}\right),
\label{eq:lang-rhn}
\end{equation}
where $\varPhi$ and $L_\alpha = (e_{L\alpha}, \nu_\alpha)^T$, $(\alpha = e, \mu, \tau)$ are Higgs and lepton weak doublets, respectively. The Yukawa coupling matrix of RH neutrinos ${N_R}_{i}$ is represented as $Y_{\alpha i}$. Here, we work on the basis that the mass matrix of the charged leptons is diagonal. The Majorana mass matrix $M_R$ is taken to be real and positive. After electroweak symmetry breaking through the SM Higgs field acquiring a vacuum expectation value $\upsilon$, the mass matrix of the neutrinos in the basis $(\nu_R^C~~ N_R)^T$ can be obtained as
\begin{equation}
 \mathcal{M} = 
 \begin{pmatrix}
  0 & m_D\\
  m_D^T & M_R
 \end{pmatrix}.
\label{eq:mass-matrix-tot}
\end{equation}

Here, $m_D = Y \upsilon$ is the Dirac mass matrix. The magnitude of $M_R$ is experimentally unconstrained. The dimension of the mass matrix $\mathcal{M}$ depends on the number of sterile neutrinos being incorporated. For $N=3,$ that means for the addition of three sterile neutrinos, it will be a symmetric matrix $6 \times 6$, which can be diagonalized by a unitary matrix of the same dimension. 
Any unitary matrix $\mathcal{U}$ of dimension $6\times6$ can diagonalize $\mathcal{M}$ such that
\begin{equation}
 \mathcal{U}^\dagger
\mathcal{M} \mathcal{U}^* = 
 \begin{pmatrix}
  \hat{m}_\nu & 0\\
  0 & \hat{M}
 \end{pmatrix},
 \label{eq:diagonalising}
\end{equation}
where $\hat{m}_\nu = \rm{Diag}(m_1, m_2, m_3)$ and 
$\hat{M} = \rm{Diag}(M_1,M_2, M_3)$ are the mass matrices of active and sterile neutrinos 
in their respective mass bases. So, the new lepton mixing matrix $\mathcal{U}$ is based on a framework where the neutral fermion spectrum comprises six mass eigenstates ($\nu_1, \nu_2, \nu_3, N_1, N_2, N_3$). We are considering the same parametrization process as we did in our previous work \cite{Das:2023aic}, so that the matrix $\mathcal{U}$ can be constructed as a product of $15$ two-dimensional rotational matrices, containing 15 rotational angles ($\theta_{ij}$) and 15 phases ($\phi_{ij}$), in a six dimensional complex space as follows:
\begin{equation}
{\mathcal{U}} = O_{56}O_{46}O_{45}O_{36}O_{26}O_{16} O_{35} O_{25}O_{15}O_{34}O_{24}O_{14}O_{23}O_{13}O_{12}.
\label{eq:U-expression}
\end{equation}
The above rotations are of the following form (exemplified by $O_{56}$):
\begin{equation}
O_{56}=
\begin{pmatrix}
1&0&0&0&0&0\\0&1&0&0&0&0 
\\0&0&1&0&0&0\\0&0&0&1&0&0\\0&0&0&0&c_{56}&s_{56}e^{-i\phi_{56}}
\\0&0&0&0&-s_{56}e^{i\phi_{56}}&c_{56}\end{pmatrix},
\label{eq:rot-matrix}
\end{equation}
where $c_{ij} = \cos \theta_{ij}$  and $s_{ij} = \sin \theta_{ij}$,  $\theta_{ij}$ are the  mixing angles and  $\phi_{ij}$ represent the CPV phases, respectively. Here, we have taken $\phi_{ij} = \delta_{ij} + \rho_{ij}$; $\delta_{ij}$ and $\rho_{ij}$ as the Dirac-type and Majorana-type phases with ($\delta_{ij},\rho_{ij} \in [0,2\pi]$).

In line with equations (\ref{eq:mass-matrix-tot}) and (\ref{eq:diagonalising}), the following relation holds
\begin{equation}
\sum_{i=1}^6 m_i \mathcal{U}^2_{ei}  = 0.
\label{eq:ee-element1}
\end{equation}
For a better understanding of the impact of both active and sterile neutrinos, let us write the matrix $\mathcal{U}$ in a block format $2\times2$ where each block will be a matrix $3\times3$:
\begin{equation}
 \mathcal{U} =
 \begin{pmatrix}
 V & R\\
  S & U
 \end{pmatrix},
\label{eq:curl-U}
\end{equation}
Now, Eq.(\ref{eq:ee-element1}) can be written as
\begin{equation}
\sum_{i=1}^3 m_i V^2_{ei} + \sum_{k=1}^3 M_k R^2_{ek} = 0.
\label{eq:ee-element}
\end{equation}
Since the entries of the matrix $\mathcal{U}$ are complex, we can make use of their real and imaginary part explicitly in Eq.(\ref{eq:ee-element}) and obtain the following:
\begin{equation}
\sum_{k=1}^3 M_k {\rm Re} R^2_{ek} =  -\sum_{i=1}^3 m_i {\rm Re} V^2_{ei},
\label{eq:linear-1}
\end{equation}
\begin{equation}
\sum_{k=1}^3 M_k {\rm Im} R^2_{ek} = -\sum_{i=1}^3 m_i {\rm Im}V^2_{ei},
\label{eq:linear-2}
\end{equation}
and with the help of the trace property of a matrix, we can write
\begin{eqnarray}
\left(\sum_i R_{i1}^2\right)M_1 + \left(\sum_i R_{i2}^2\right)M_2 + \left(\sum_i R_{i3}^2\right)M_3 &=&\nonumber \\
-\left(\left(\sum_i V_{i1}^2\right)m_1 + \left(\sum_i V_{i2}^2\right)m_2 + \left(\sum_i V_{i3}^2\right)m_3\right).
\label{eq:linear-3}
\end{eqnarray}
These three linear equations (\ref{eq:linear-1}-\ref{eq:linear-3}) enable us to obtain the following analytical expressions for the added sterile states:
\begin{equation}
    M_1=\frac{d_1(b_3 c_2 - b_2 c_3) + d_2(b_1 c_3 - b_3 c_1) + d_3(b_2 c_1 - b_1 c_2)}{a_1(b_2c_3 - b_3c_2) - a_2(b_1c_3 - b_3c_1) + a_3(b_1c_2 - b_2c_1)},
    \label{eq:sol-mass1}
\end{equation}
 \begin{equation}
    M_2=\frac{d_1(a_2 c_3 - a_3 c_2) + d_2(a_3 c_1 - a_1 c_3) + d_3(a_1 c_2 - a_2 c_1)}{a_1(b_2c_3 - b_3c_2) - a_2(b_1c_3 - b_3c_1) + a_3(b_1c_2 - b_2c_1)},
    \label{eq:sol-mass2}
\end{equation}
 \begin{equation}
   M_3=\frac{d_1(a_3 b_2 - a_2 b_3) + d_2(a_1 b_3 - a_3 b_1) + d_3(a_2 b_1 - a_1 b_2)}{a_1(b_2c_3 - b_3c_2) - a_2(b_1c_3 - b_3c_1) + a_3(b_1c_2 - b_2c_1)},
   \label{eq:sol-mass3}
\end{equation}
\vspace{0.5pt}
where 
 \begin{equation*}
    \begin{split}
         a_1={\rm Re} R_{e1}^2,\quad b_1={\rm Re} R_{e2}^2,\quad c_1= {\rm Re}R_{e3}^2,\quad d_1=\sum_{i=1}^3 m_i {\rm Re} V^2_{ei},\\
          a_2={\rm Im} R_{e1}^2,\quad b_2={\rm Im} R_{e2}^2,\quad c_2={\rm Im} R_{e3}^2,\quad d_2=\sum_{i=1}^3 m_i {\rm Im} V^2_{ei},\\
         a_3=\sum_{i=1}^3 R_{i1}^2,\;\quad b_3=\sum_{i=1}^3 R_{i2}^2,\quad c_3=\sum_{i=1}^3 R_{i3}^2,\quad d_3=\sum_{i,j=1}^3 V_{ij}^2 m_j.
    \end{split}
    \label{eq:abbrieviation}
\end{equation*}

\section{Neutrinoless double beta decay in $\chi$EFT framework }
\label{sec:ovbb-eft}
We intend to explore the impact of the added sterile neutrinos on the half-life of the $0\nu \beta \beta$ decay.
Assuming the decay process to be mediated by both active and sterile neutrinos, the inverse of the half-life period can be computed through
\begin{equation}
     \left(\thalf\right)^{-1} = G_{01}\, g_A^4 \left| \frac{V_{ud}^2}{m_e}\right|^2 \left| \sum_{i=1}^6 \, m_i\,\mathcal{U}_{ei}^2\,\amp(m_i) \right|^2,
    \label{half-life1}
\end{equation}
 where $G_{01}$ is a phase-space factor that has a particular value for the specific isotopes considered in the decay process. For  $^{136}\text{Xe}$ , the value of $G_{01} \simeq 1.4\cdot 10^{-14}\text{ y}^{-1}$ \cite{Kotila:2012zza}. $g_A \simeq 1.27$ is the nucleon axial coupling and $V_{ud}\simeq 0.97$ is the up-down CKM matrix element \cite{ParticleDataGroup:2024cfk}. Here $i$ runs over all neutrino mass eigenstates (both active and sterile). 
 The amplitudes $\amp(m_i)$, which incorporate all contributions of hadronic and nuclear physics, are the key quantities to examine the impact of active and sterile neutrinos in a distinct way. Evaluation of $\amp(m_i)$ requires the $\chi$EFT approach. For clarity, hereafter we denote the mass states of sterile neutrinos as $M_i$ with $(i=1,2,3)$ instead of the previous notation $m_i$ with $(i=4,5,6)$.
So far, the underlying mechanism of the $0\nu \beta \beta$ decay is concerned, it can be regulated by the exchange of either light active neutrinos or sterile neutrinos or both \cite{Blennow:2010th, Mitra:2011qr, Asaka:2016zib, Faessler:2014kka}.  When the contributions from both are taken into account, the $0\nu \beta \beta$ decay rate is found to be affected in accordance with the mass range of sterile neutrinos. For heavier mass regimes of sterile neutrinos such as $>1~\rm{GeV}$, the $0\nu \beta \beta$ decay process will receive dominant contribution from the light active neutrinos, while for the sub-GeV mass regimes of sterile neutrinos, one may observe enhancement or supression depending on other parameters like mixing angles and $CP$ violating phases as pointed out in references \cite{Dekens:2024hlz, Dekens:2023iyc, Huang:2019qvq}. Besides these particle physics aspects, the accurate prediction of the $0\nu \beta \beta$ half-life is also dependent on the nuclear physics aspect arising through the nuclear matrix elements (NME). Conventionally, these NMEs are calculated within specific nuclear frameworks like shell model \cite{Caurier:2007qn, Brown:2001zz}, quasi-particle random phase approximation \cite{Pantis:1996py, Bobyk:2000dw}, interacting boson model \cite{Arima:1981hp, Barea:2009zza}, energy density functional model \cite{Rodriguez:2010mn,LopezVaquero:2013yji}, etc. Due to the complex nature of nuclear interactions, their outcomes differ frequently and subsequently introduce significant theoretical uncertainties. These uncertainties can be minimized with the aid of the $\chi$EFT, which only focuses on the most suitable low-energy physics inside the nucleus \cite{Weinberg:1990rz, Weinberg:1991um, Ordonez:1992xp}. In the $\chi$EFT framework, the interactions are generally expressed in terms of protons, neutrons, and pions, employing a set of operators constructed in line with the intrinsic symmetry of the theory, thereby facilitating the precise control over the uncertainties. Further, a different mass range of sterile neutrinos is tailored essentially in the $\chi$EFT framework \cite{Cirigliano:2017tvr, Brase:2021uny}. $\chi$EFT integrates out the effect of heavy sterile states by portraying them in terms of short-distance interactions, whereas the effect of lighter sterile states is treated in terms of long-distance interactions. In this way, a consistent connection between the NME evaluation and the LNV interactions across all mass scales accountable for the $0\nu \beta \beta$ decay can be established through $\chi$EFT, and this predicts the $0\nu \beta \beta$ half-life more reliably.

For an efficient $\chi$EFT approach, the identification of a separating scale between the high-energy and low-energy dynamics is a decisive factor. In the case of $0\nu \beta \beta$ decay, the nuclear momentum exchange occurs at the scale around $\sim 100 ~\rm{MeV}$. Further, the high-scale LNV interactions and the low-scale hadronic degrees of freedom can be aligned around the QCD scale, i.e., $\Lambda_{\chi} \sim 1~ \rm{GeV}$. This scale $\Lambda_{\chi}$ is also known as the chiral-symmetry breaking scale below which the nucleons and pions become relevant degrees of freedom. The loop momentum scale pertinent for the neutrinos exchanged in the $0\nu \beta \beta$ decay can have a range from below the nuclear momentum scale $\sim 100 ~\rm{MeV}$ up to the cut-off scale $\Lambda_{\chi}$ and above. So, we will investigate the behavior of the mediating neutrinos in these momentum regions and their contribution to the decay amplitude. As pointed out in the references \cite{Beneke:1997zp, Machleidt:2011zz}, the following contributions are expected if we consider one scale at a time:
\begin{itemize}
    \item for the momentum scale $k_0\sim |\mathbf {k}| \sim \Lambda_\chi$, $\chi$EFT will consider the contribution of Hard neutrinos through short-distance interactions.
    \item for the momentum scale $k_0\sim |\mathbf {k}| \sim m_\pi$, Soft neutrinos are found to play the role of leading contributor to the loop diagrams involving nucleons and pions.
    \item For the momentum scale $k_0\sim |\mathbf {k}|^2/m_N \sim k_F^2/m_N$, where $k_F$ is the Fermi momentum $\sim 100 ~\rm{MeV}$ and $m_N$ is the nucleon mass, Potential neutrinos come to the scenario through long-distance interactions.
    \item If the momentum scales as $k_0\sim |\mathbf {k}| \sim k_F^2/m_N$, Ultrasoft neutrinos having coupling with the whole nucleus instead of individual nucleons, are of primary interest.
\end{itemize}
Based on the above momentum distribution, the $0\nu\beta\beta$ decay half-life can be written in a revised way explicitly as follows:
 \begin{equation}
\begin{split}
\left(\thalf\right)^{-1}
=\, &G_{01}\, g_A^4 \left|\frac{V_{ud}^2}{m_e}\right|^2
\Bigg|
\left(m_1\,\mathcal{U}_{e1}^2 + m_2\,\mathcal{U}_{e2}^2 + m_3\,\mathcal{U}_{e3}^2\right)\,\mathcal{A}(0)
\\[3pt]
&+ \left(M_1\,\mathcal{U}_{e4}^2\,\mathcal{A}(M_1)
+ M_2\,\mathcal{U}_{e5}^2\,\mathcal{A}(M_2)
+ M_3\,\mathcal{U}_{e6}^2\,\mathcal{A}(M_3)\right)
\Bigg|^2 .
\end{split}
\label{half-life}
\end{equation}  
 In the context of active neutrino exchange, the amplitude is defined as $\amp(0)$. As their masses lie in the sub-eV mass range, we can effectively consider $m_i = 0$ for the evaluation of the amplitude $\amp (m_i)$. However, we will have a detailed exploration of sterile neutrino masses and their contribution to the decay amplitude in the following subsection. 
 
The light active neutrinos, having masses much less than that of the nuclear momentum scale, are essentially subjected to the long-distance contributions. However, as per some recent studies presented in references \cite{Cirigliano:2019vdj, Cirigliano:2018hja}, the $\chi$EFT allows the momentum integral over the virtual neutrino propagator to probe higher momenta larger than the nuclear scale. These effects from the hard neutrino region are incorporated in the short-distance part.
 As discussed in references \cite{deVries:2024rfh, Dekens:2024hlz}, the amplitude $\amp(0)$ can be defined as a sum of long-distance and short-distance contributions as follows.
 \begin{equation}
     \amp(0)= \amp^{\rm (ld)}(0)+\amp^{\rm (sd)}(0). 
    \label{amp-active}
 \end{equation}
The long-distance part $\amp^{\rm (ld)}$ and short-distance part $\amp^{\rm (sd)}$ in terms of NMEs are given by,
\begin{align}
\amp^{\rm (ld)}(0)
  &= -\,\mathcal{M}(0)
   \;\equiv\;
   -\,\frac{\mathcal{M}_F}{g_A^2}
   + \mathcal{M}_{GT}
   + \mathcal{M}_T,  \label{amp-active-ld}\\[4pt]
\amp^{\rm (sd)}(0)
  &= -\,2\, g_\nu^{NN}\, m_\pi^2\,
     \frac{\mathcal{M}_{F,\mathrm{sd}}}{g_A^2}.
      \label{amp-active-sd}
\end{align}
 Here $\mathcal{M}_{i}$ are the total Fermi (F), Gamow-Teller (GT), and tensor (T)  nuclear matrix elements (NMEs) having minuscule dependence on the mass of the exchanged active neutrinos. $\mathcal{M}_{F,sd}$ is normalized to $\mathcal{O}(1)$ and drives the short-range NME, and $g_\nu^{NN}$ is a QCD matrix element whose value will be defined later.
 
Now, in the context of sterile neutrino contribution, following \cite{deVries:2024rfh, Dekens:2024hlz}, the mass dependence of the amplitude can be demarcated through the following effective formula connecting all three regions based on the $\chi$EFT approach:
\begin{eqnarray}
\mathcal{A} (M_i) = \begin{cases}
\mathcal{A}^{\rm (ld,<)}(M_i)+\mathcal{A}^{\text{(sd)}}(M_i)+\mathcal{A}^{\text{(usoft)}}(M_i) \,,&  M_i <  100 \text{ MeV}\,, \\
\mathcal{A}^{\rm (ld)}(M_i)+\mathcal{A}^{\text{(sd)}}(M_i)\,,& 100 \text{ MeV} \le M_i <  2 \text{ GeV}\,,\\
\mathcal{A}^{\text{(9)}}(M_i)\,,& {\rm }\,  2 \text{ GeV} \le M_i \,,
\end{cases}
\label{mass-dependent-amp}
\end{eqnarray}
The relevant expressions required for the computation of  $\amp(M_i)$ depending on the region in which the sterile neutrino mass belongs are explicitly given in Appendix \ref{amp-formula}.

\subsection{Effective neutrino mass of $0\nu \beta \beta$ decay}
There is another important quantity affecting the sensitivity of the $0\nu \beta \beta$ decay process. It is known as the effective neutrino mass ($m_{ee}$). So far as the decay process is considered to be executed by exchange of SM active neutrinos, this $m_{ee}$ can be expressed as a combination of the $i^{th}$ mass eigenstate of the neutrinos multiplied by the square of the electron-flavor mixing matrix element. But when we are considering contributions from sterile neutrinos, the effective neutrino mass will get additional contributions from the active-sterile mixing, $ CP$-violating phases of added sterile states. Nevertheless, the effective neutrino mass $m_{ee}$ is connected to the $0\nu \beta \beta$ half-life in an inverse proportion, i.e., $|m_{ee}|^2 \propto 1/\thalf$ in the high-scale seesaw limit. Since we are considering different mass ranges in the low-scale seesaw limit, it will be reasonable to define the effective neutrino mass in connection with the half-life as follows \cite{Cirigliano:2018yza}:
\begin{equation}
    m_{ee}=\frac{m_e}{g_A^2 |V_{ud}|^2 \mathcal{M}(0) G_{01}^{1/2}} \left( \thalf \right)^{-1/2}.
    \label{mee}
\end{equation}
\section{Results}\label{result-eft}
First, we will define the parameter space using the analytical expressions for the masses of the sterile neutrinos given in 
equations (\ref{eq:linear-1} - \ref{eq:linear-3}). They encapsulate 15 mixing angles, 15 phases, and the light neutrino mass states. The 15 mixing angles comprise 3 active-active, 3 sterile-sterile, and 9 active-sterile mixing angles. Out of these, we are disregarding the sterile sector mixing so that $\theta_{45}=\theta_{46}=\theta_{56}=0$. The three mixing angles from the active sector ($\theta_{12},\theta_{13}$ and $\theta_{23}$) have their experimentally observed values \cite{Esteban:2024eli}. This leaves behind only nine active-sterile mixing angles ($\theta_{14}, \theta_{24}, \theta_{34}, \theta_{15}$, 
$\theta_{25},\theta_{35}, \theta_{16}, \theta_{26}, \theta_{36}$) as free parameters. 
Multiple factors imply minimal values for these active-sterile mixing angles. Some of them include the experimental testability of the low-scale seesaw \cite{Abdullahi:2022jlv, Bolton:2019pcu}, validation of the oscillation, and the precision electro-weak data \cite{Xing:2007zj, Xing:2009in}. Motivated by these factors, the following set of values for the active-sterile mixing angles have considered in Ref. \cite{Das:2023aic}, where a mass scale between the sub-keV to TeV range for sterile neutrinos was obtained.
\begin{itemize}
 \item Set 1: $\theta_{1j}\approx10^{-5}$, $\theta_{2j}\approx10^{-4}$, $\theta_{3j}\approx10^{-3}$,
 \item Set 2:  $\theta_{1j}\approx10^{-4}$, $\theta_{2j}\approx10^{-3}$, $\theta_{3j}\approx10^{-2}$, 
 \item Set 3: $\theta_{1j}\approx10^{-3}$, $\theta_{2j}\approx10^{-2}$, $\theta_{3j}\approx10^{-1}$.
\end{itemize}
We also implement these values for our subsequent calculations. 
\begin{figure}
    \centering
    \subcaptionbox{\label{fig:sub13}}{\includegraphics[width=4.8cm,height=5cm]{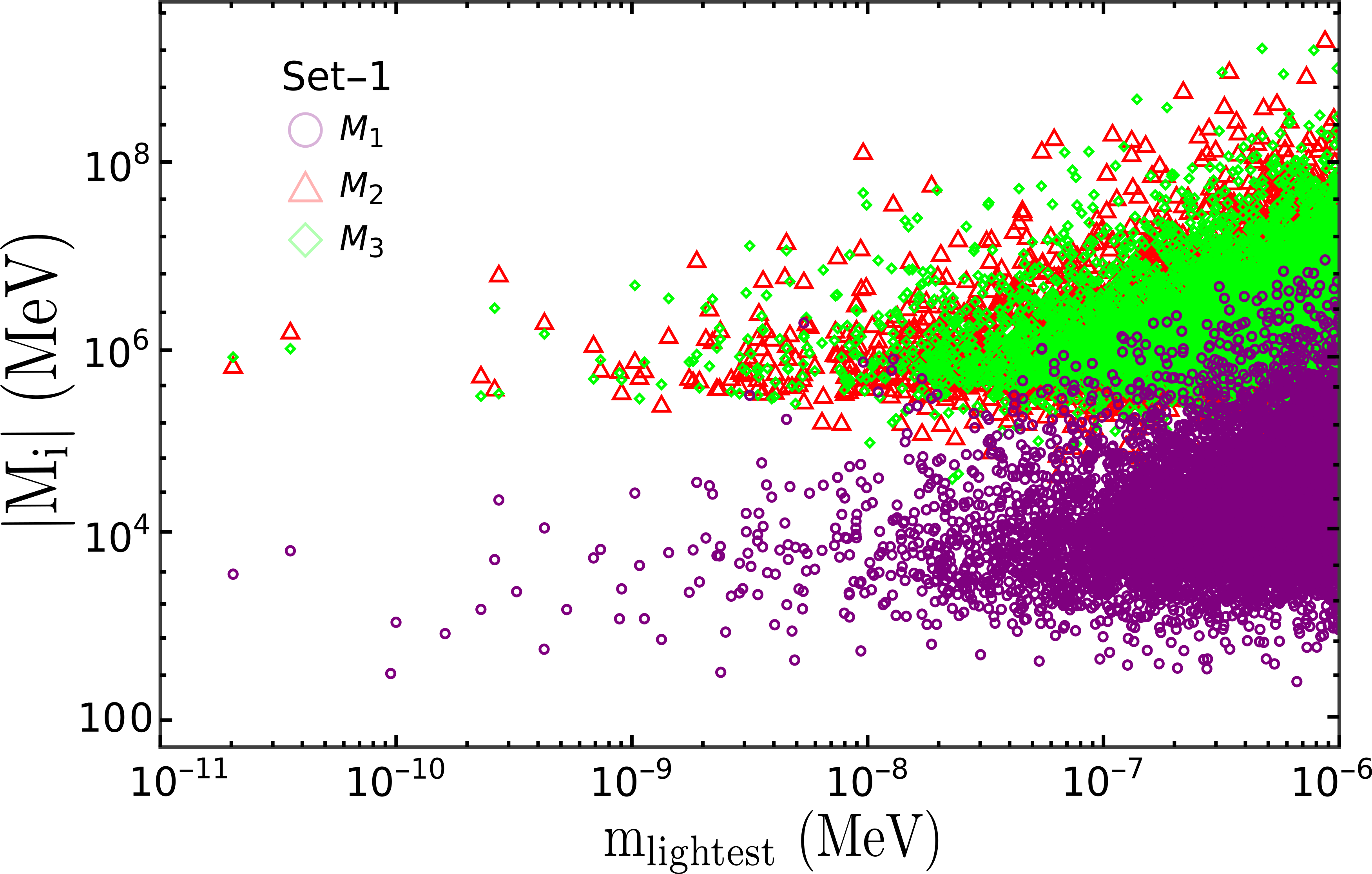}}
    \hspace{0.2cm}
    \subcaptionbox{\label{fig:sub14}}{\includegraphics[width=4.8cm,height=5cm]{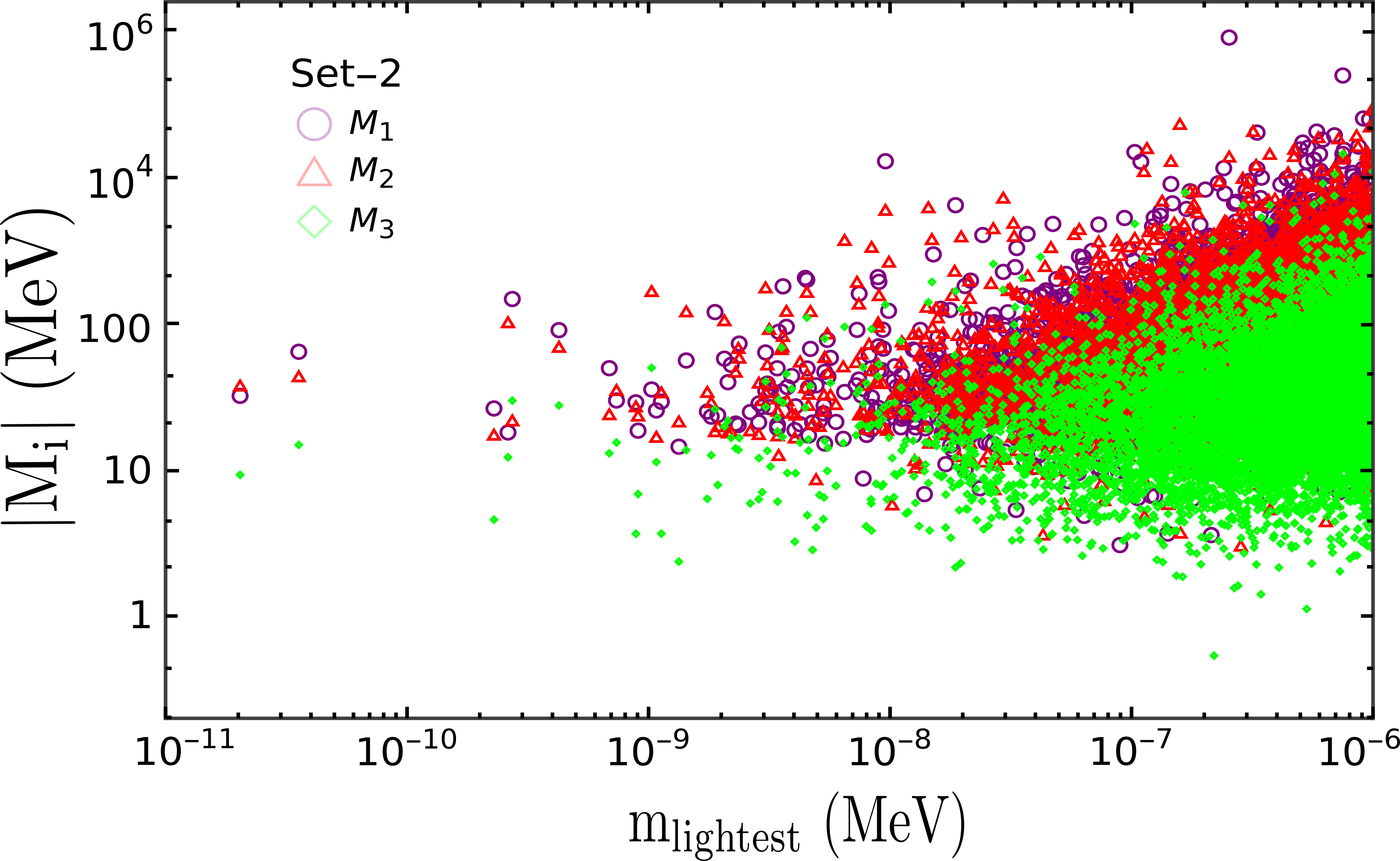}}
    \hspace{0.2cm}
    \subcaptionbox{\label{fig:sub15}}{\includegraphics[width=4.8cm,height=5cm]{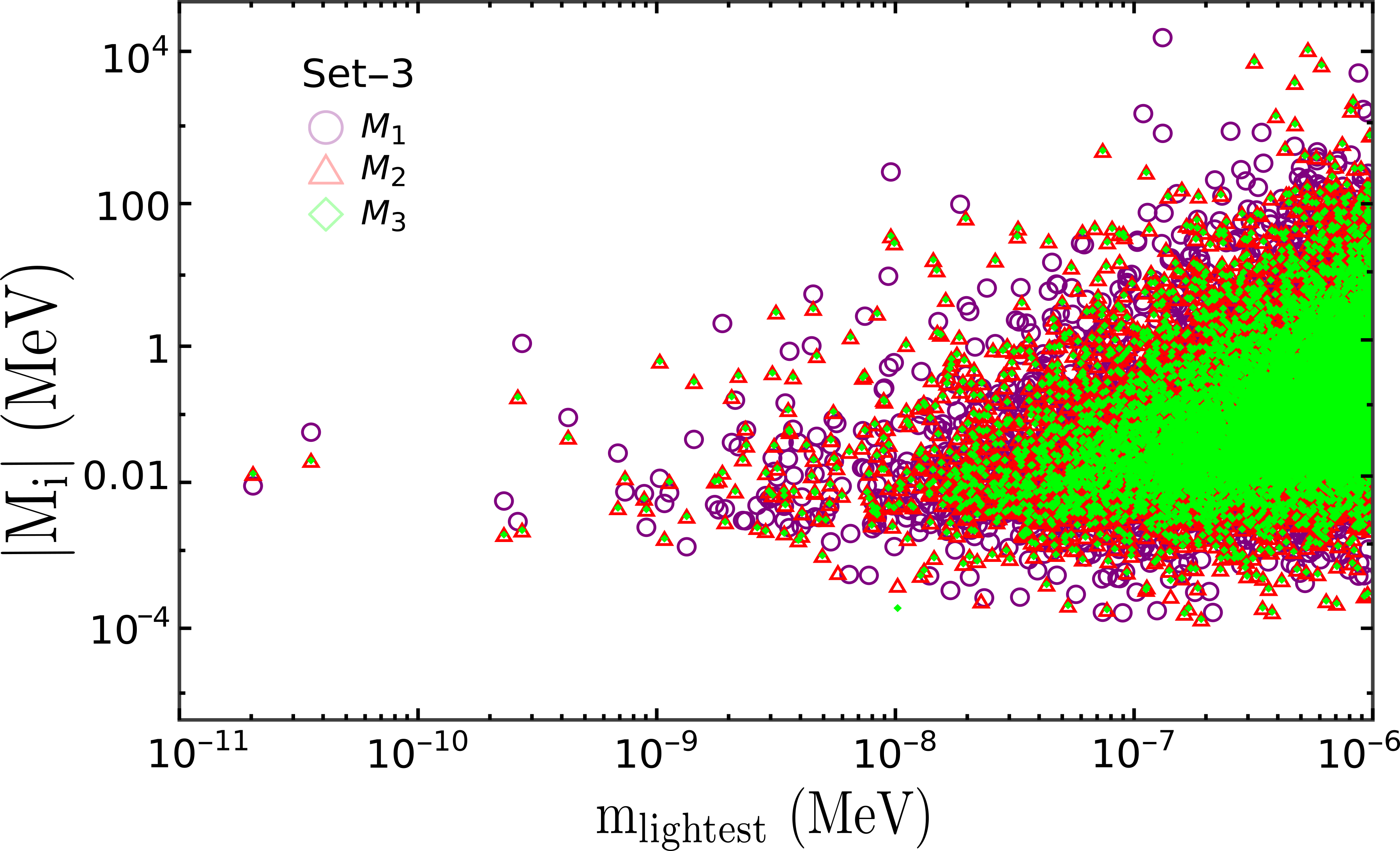}} \\
    \caption{Figure depicting the possible mass range of the sterile neutrinos. These plots are obtained using the analytical expressions for the sterile neutrinos shown in Eqs. (\ref{eq:linear-1}-\ref{eq:linear-3}) as a function of $m_{lightest}$.  }
    \label{fig-sterilemass}
\end{figure}
Similarly, out of 15 phases, the five phases $\phi_{12},\phi_{13}, \phi_{14}, \phi_{15}$ and $\phi_{16}$ are selected as prime contributor according to the equations (\ref{eq:linear-1} - \ref{eq:linear-3}), and the contribution of the rest of the phases can be ignored for simplicity. As pointed out in 
Ref. \cite{Das:2023aic}, these five phases can have an independent variation from $0$ to $2\pi$, but with the constraint $\phi_{14} \ne \phi_{15} \ne \phi_{16} \ne \frac{n\pi}{2},\; n=0,1,2,3, \cdots$, owing to the expressions for $M_1,~ M_2~ \rm{and}~M_3$. If this constraint is not followed, the expressions diverge to infinity.  Among the active neutrino mass states, two heavier masses can be expressed in terms of the lightest one among them according to the normal hierarchy (NH) or inverted hierarchy  (IH) condition. So for NH: $m_3>m_2>m_1(m_{\rm lightest})$ with $m_2 = \sqrt{m_1^2 + \Delta m_{21}^2}$ and $m_3 = \sqrt{m_1^2 + \Delta m_{31}^2}$. For IH: $m_2>m_1>m_3(m_{\rm lightest})$ with $m_1 = \sqrt{m_3^2 -\Delta m_{32}^2-\Delta m_{21}^2}$ and $m_2 = \sqrt{m_3^2 - \Delta m_{32}^2}$.

Having defined the parameter space under consideration, we now proceed to evaluate the $0\nu \beta \beta$ half-life using Eq. (\ref{half-life}) and the effective neutrino mass using Eq. (\ref{mee}). We have three regions based on the $\chi$EFT defined in the appendix \ref{amp-formula}, so we will evaluate $\thalf$ corresponding to each region. Our selected set of mixing angles and $CP$ violating phases are being employed as input in the expression for $\left(\thalf \right) ^{-1}$ through the parametric expressions of the mixing matrix elements $\mathcal{U}_{ei}$ (obtained through the equations \ref{eq:U-expression} and \ref{eq:rot-matrix}) and through the analytical expressions for the masses of the sterile neutrinos (given in equations (\ref{eq:linear-1} - \ref{eq:linear-3})). We use Eq. (\ref{amp-active}) to incorporate the light active neutrino contribution for the amplitude $\amp(0)$ in $\left(\thalf \right) ^{-1}$  for all cases, as there is no mass dependency. However, the incorporation of the sterile neutrino contribution to the amplitude through $\amp(M_i)$ in the $\left(\thalf \right) ^{-1}$ expression requires the selection of an appropriate amplitude expression. The amplitude expressions contain the combination of NMEs and hadronic matrix elements. The values of different NMEs computed using the nuclear shell model for the isotope ${}^{136}\rm{Xe}$, are given in the tables- (\ref{tab:NME}, \ref{tab:NME1}, and \ref{tab:usoftNMEforXe}). The explicit expression of the hadronic matrix element is given in Eq. (\ref{eq:gnu_int}).

Fig. (\ref{fig-sterilemass}) shows the overall mass range of the sterile neutrinos according to the mixing angle values (defined in sets -1,2, and 3) and the constraints on $ CP$-violating phases. These plots are obtained using the analytical expressions for the masses of the sterile neutrinos shown in 
equations (\ref{eq:linear-1} -  \ref{eq:linear-3}) as a function of $m_{lightest}$ (that means $m_1$ for NH and $m_3$ for IH). However, the mass range for each sterile neutrino is found to lie in a comparable range in both NH and IH. The different mass range of sterile neutrinos, as can be seen from Fig. (\ref{fig-sterilemass}), implies the possible mass regime from which $\amp(M_i)$ will get contributions. For instance, in the case of set-1, all the sterile neutrinos lie above $100$ MeV, which implies that $\amp(M_i)$ will get input from regions 1 and 2 in the calculation of $\thalf$ and $m_{ee}$. On the other hand, for sets 2 and 3, where the mass of all the sterile neutrinos span $(1-10^6)$ MeV and $(10^{-4}-10^4)$ MeV, respectively, $\amp(M_i)$ will get contributions from all three regions as per Eq. (\ref{mass-dependent-amp}).

\begin{figure}[htbp!]
    \centering
    \subcaptionbox{\label{fig:sub1}}{\includegraphics[width=5cm,height=5.2cm]{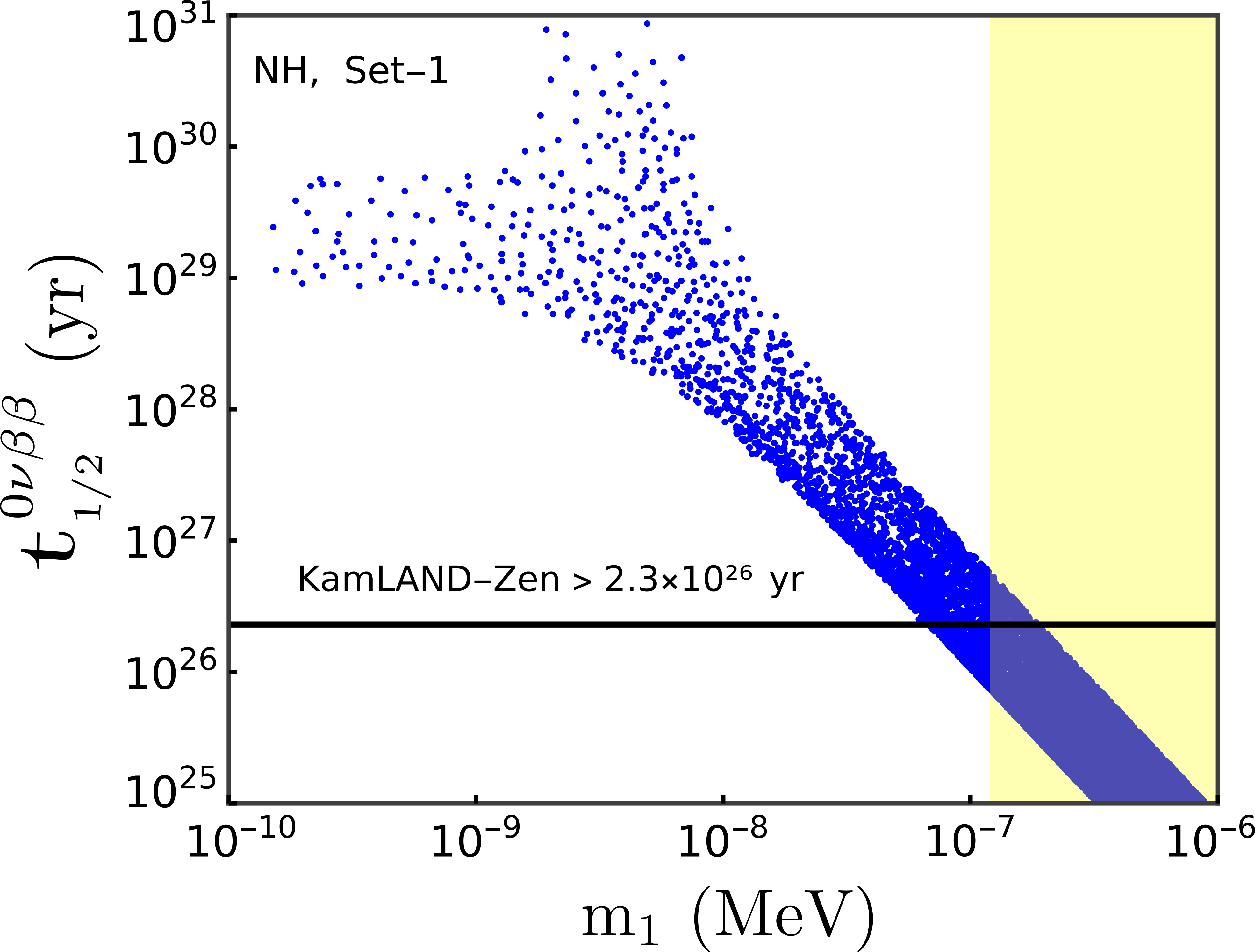}}
    \hspace{0.2cm}
    \subcaptionbox{\label{fig:sub2}}{\includegraphics[width=5cm,height=5.2cm]{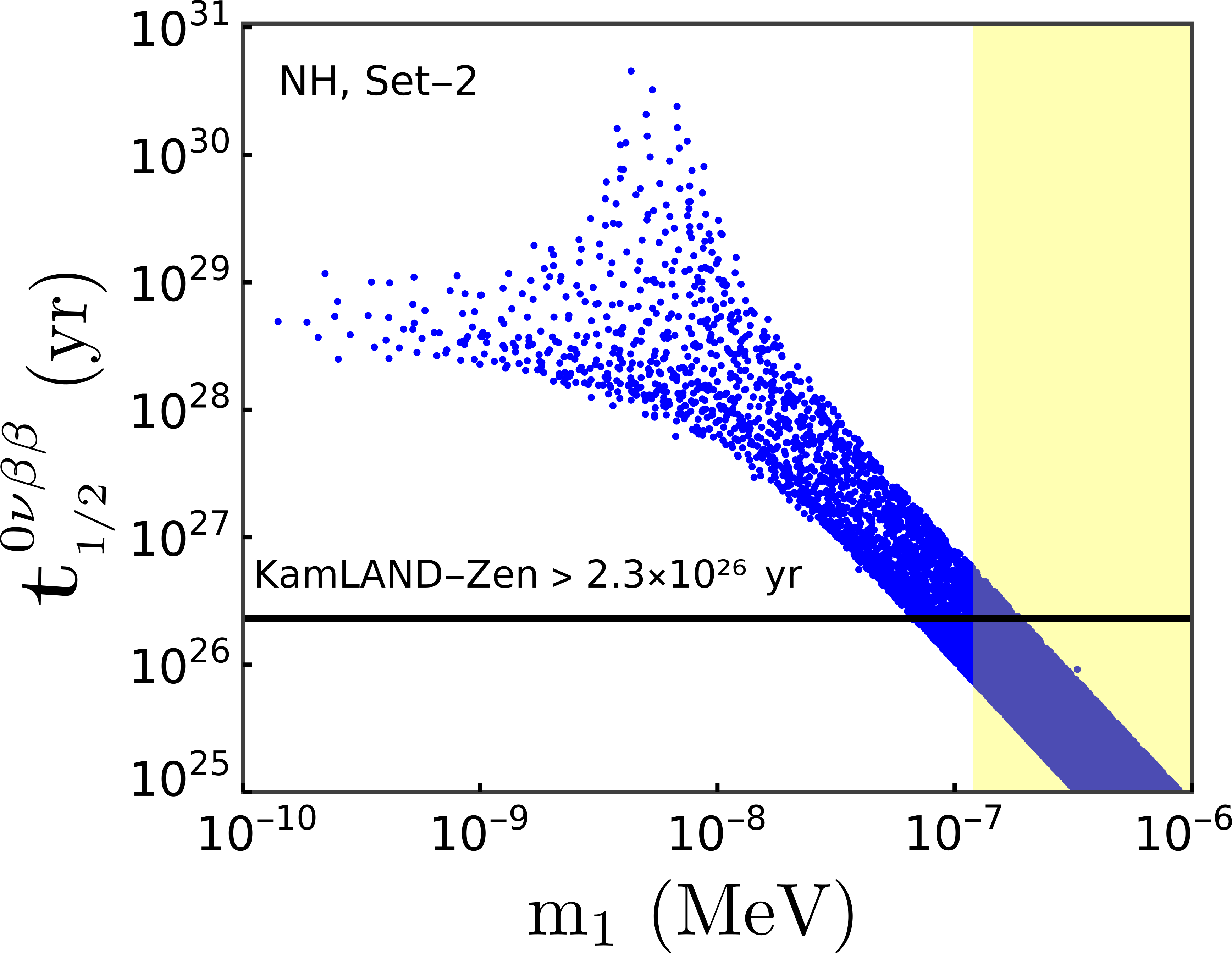}}
    \hspace{0.2cm}
    \subcaptionbox{\label{fig:sub3}}{\includegraphics[width=5cm,height=5.2cm]{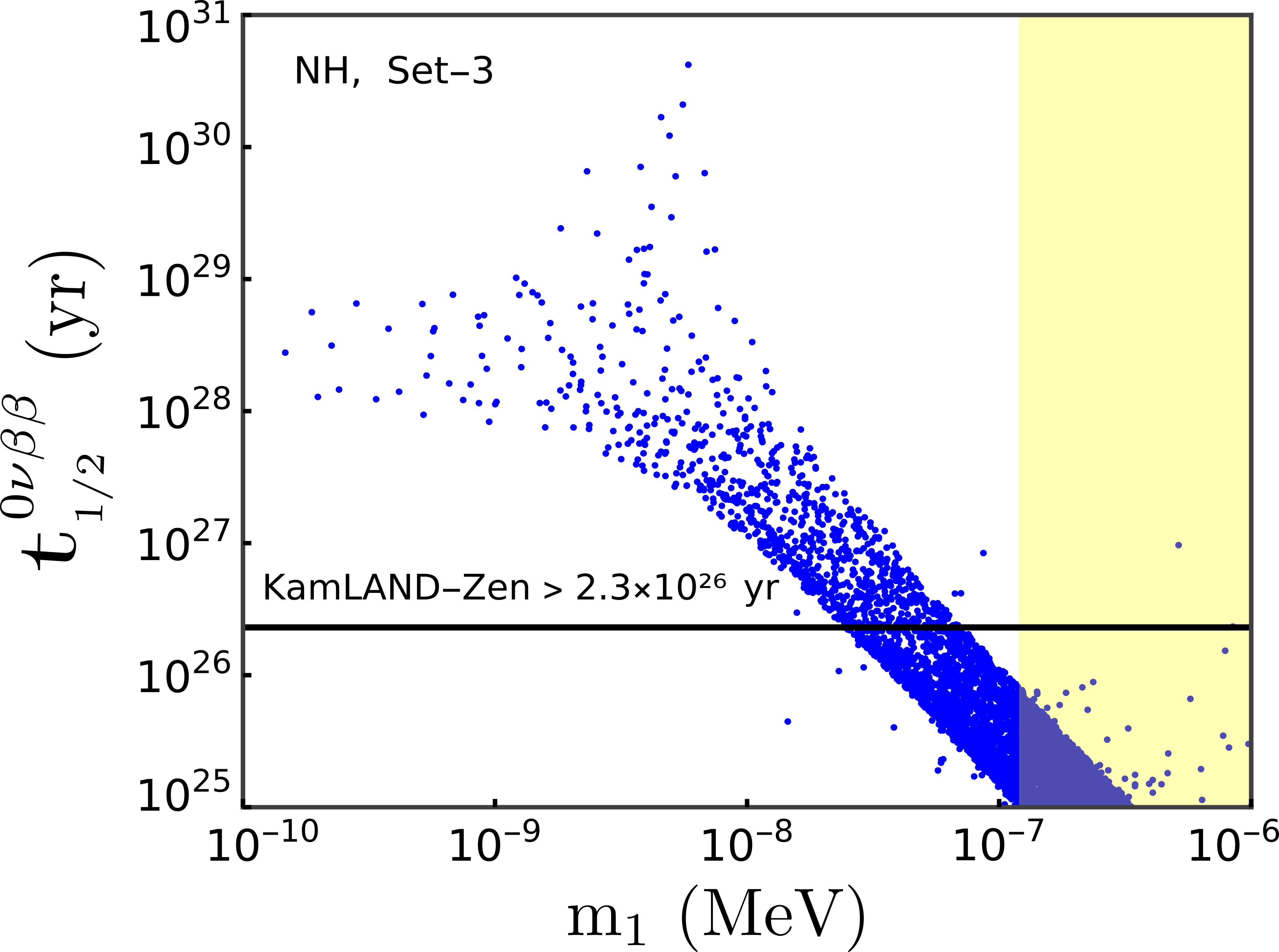}} \\
    \subcaptionbox{\label{fig:sub4}}{\includegraphics[width=5cm,height=5.2cm]{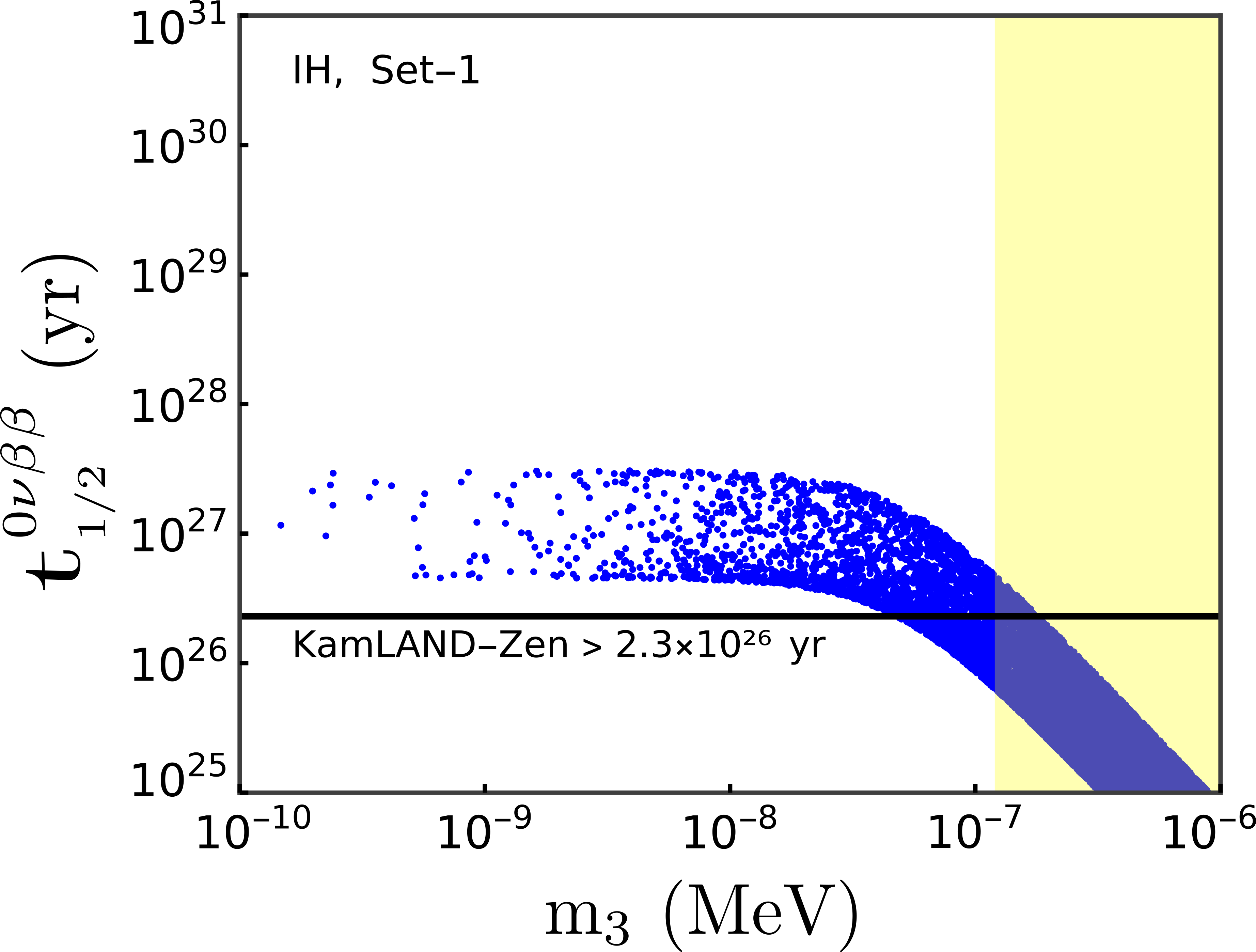}}
    \hspace{0.2cm}
    \subcaptionbox{\label{fig:sub5}}{\includegraphics[width=5cm,height=5.2cm]{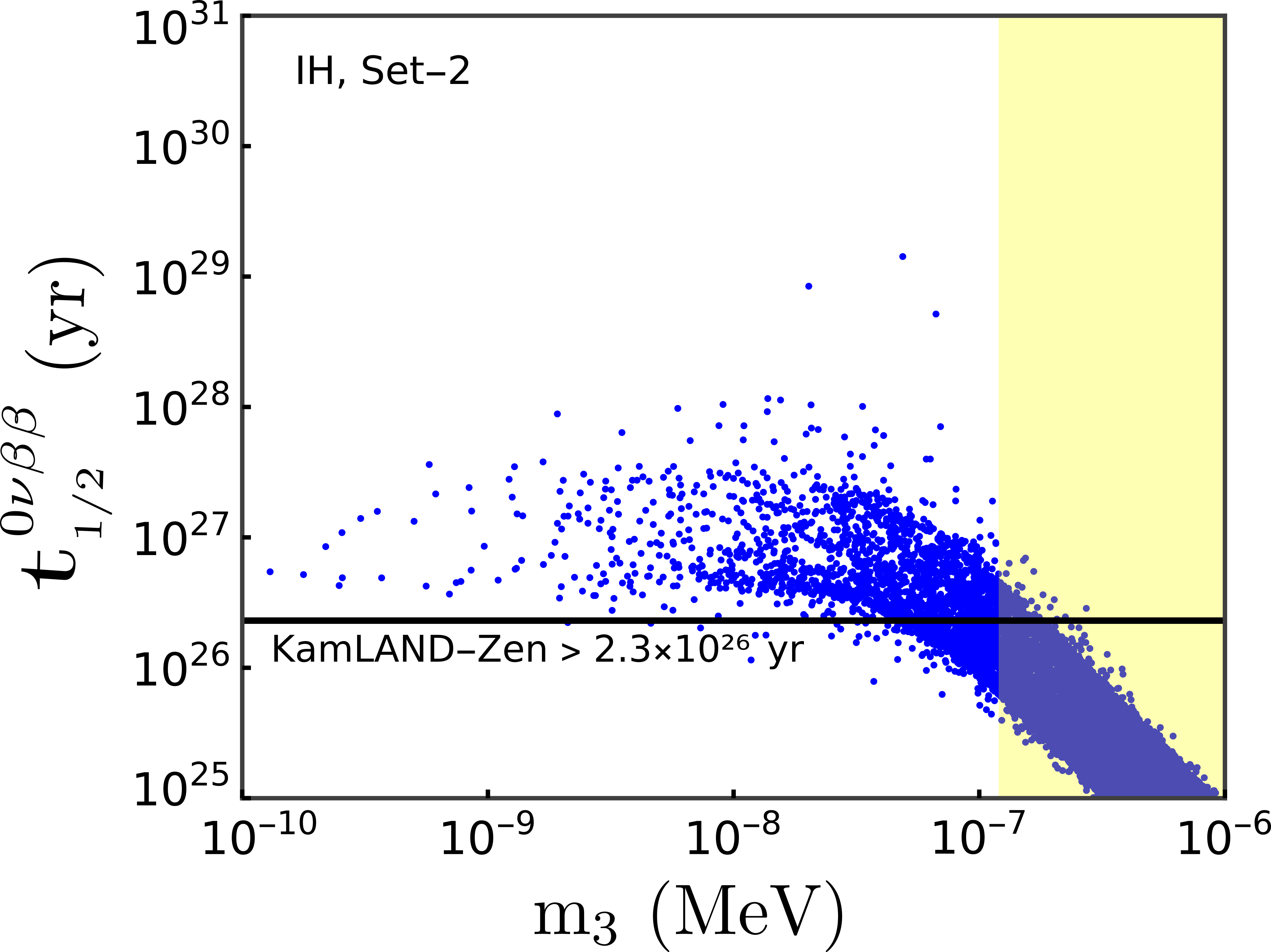}}
    \hspace{0.2cm}
    \subcaptionbox{\label{fig:sub6}}{\includegraphics[width=5cm,height=5.2cm]{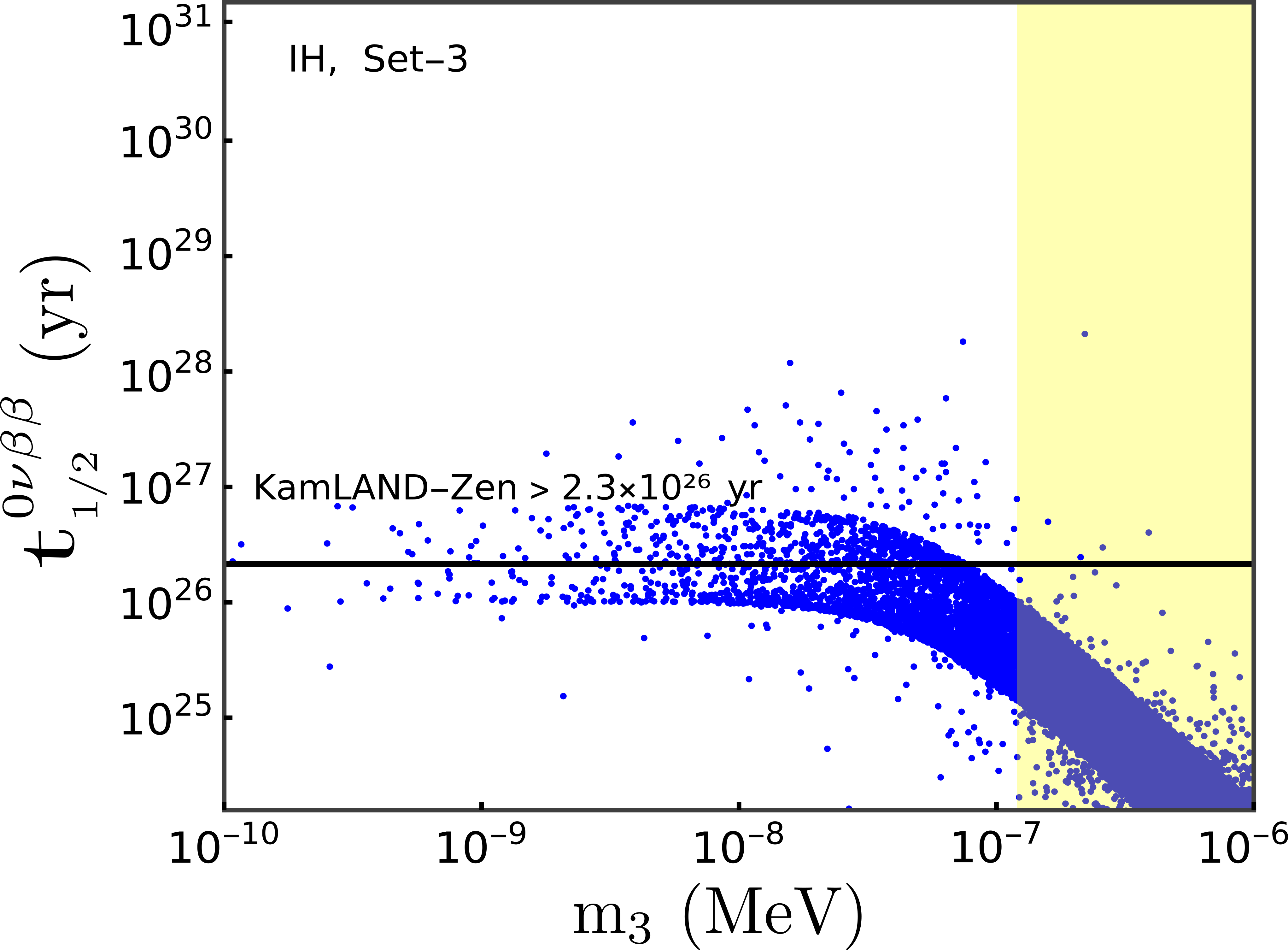}}
    \caption{Figure depicting the variation of $0\nu \beta \beta$ decay half-life of ${}^{136}\rm{Xe}$ isotope against the lightest neutrino mass state $m_1$ ($m_3$). The upper (lower) panel is for NH(IH) of light active neutrinos. The solid black line represents the current experimental bound on $\thalf$ from KamLAND-Zen. The light yellow region corresponds to the cosmologically disfavored region where the sum of light neutrino mass states exceeds $0.12\times10^{-6}$ MeV. The half-life value in case of set-1 (the extreme left plots) includes contributions from sterile neutrinos in the mass range $(100 - 10^8)$ MeV. For sets - 2 and 3, the half-life values include sterile neutrinos in the mass range $(1-10^6)$ MeV and $(10^{-4} - 10^{4})$ MeV, respectively (shown in middle and extreme right plots).}
    \label{fig-thalf}
\end{figure}

\begin{figure}
    \centering
    \subcaptionbox{\label{fig:sub7}}{\includegraphics[width=5cm,height=5.2cm]{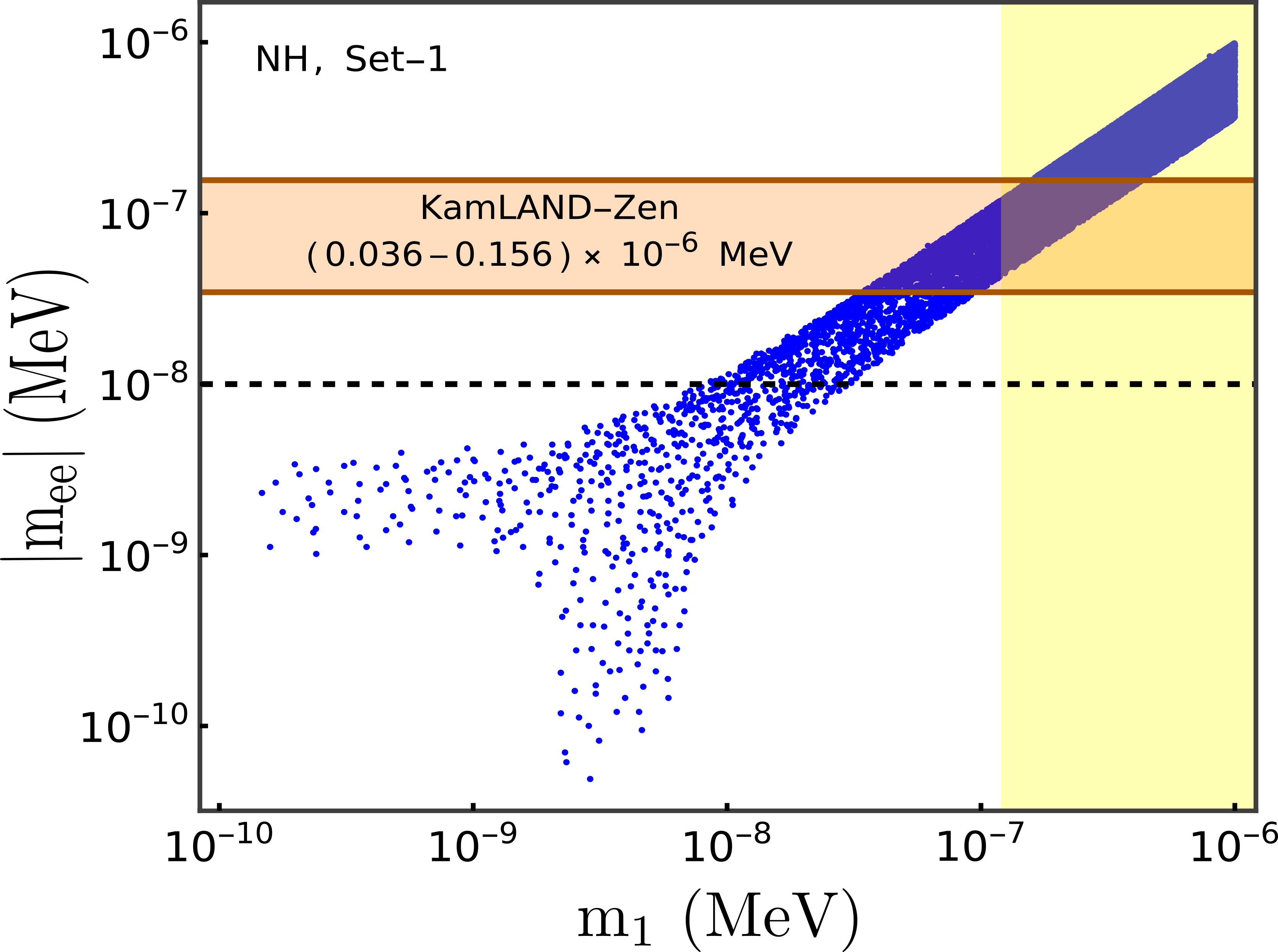}}
    \hspace{0.2cm}
    \subcaptionbox{\label{fig:sub8}}{\includegraphics[width=5cm,height=5.2cm]{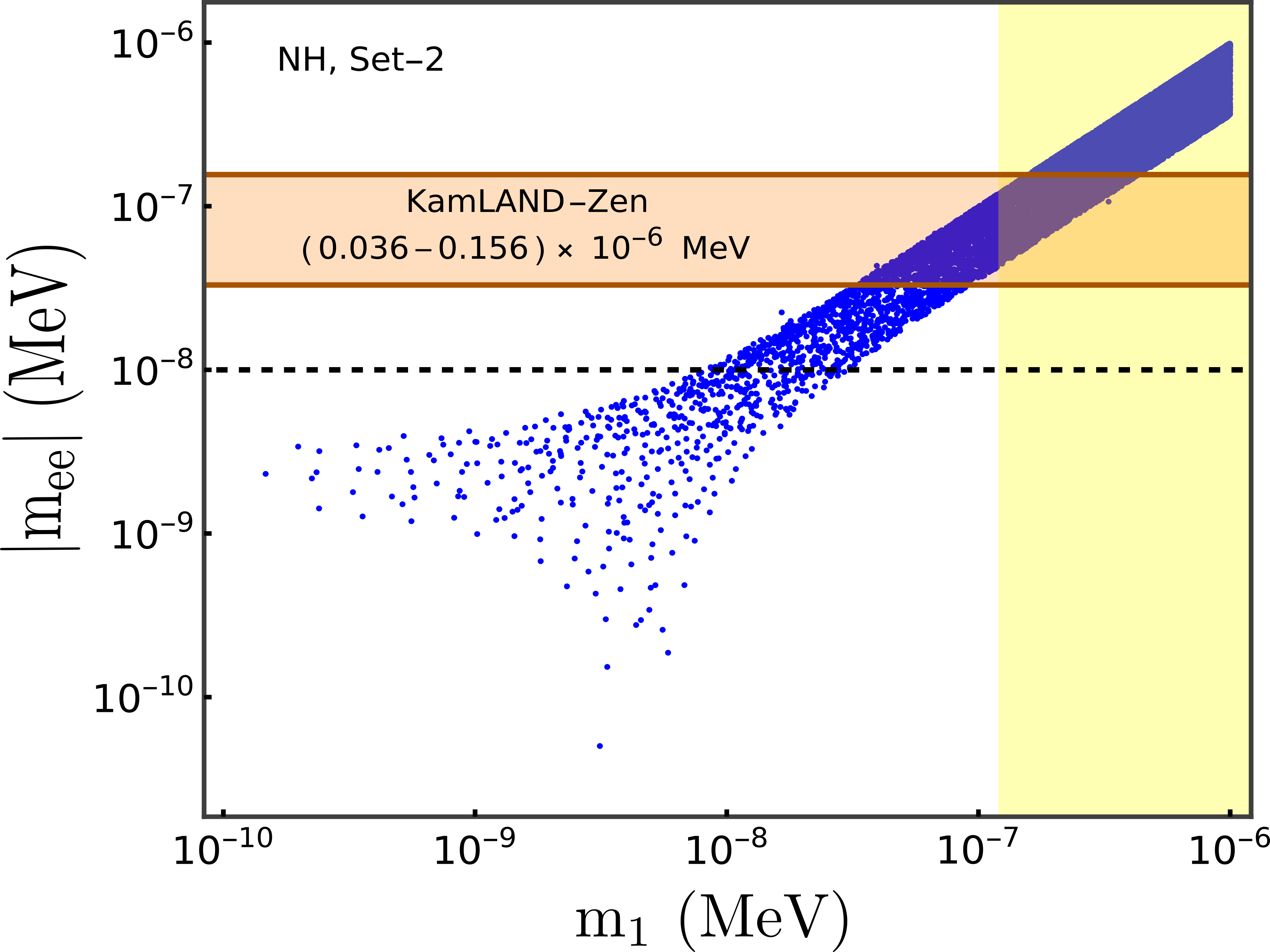}}
    \hspace{0.2cm}
    \subcaptionbox{\label{fig:sub9}}{\includegraphics[width=5cm,height=5.2cm]{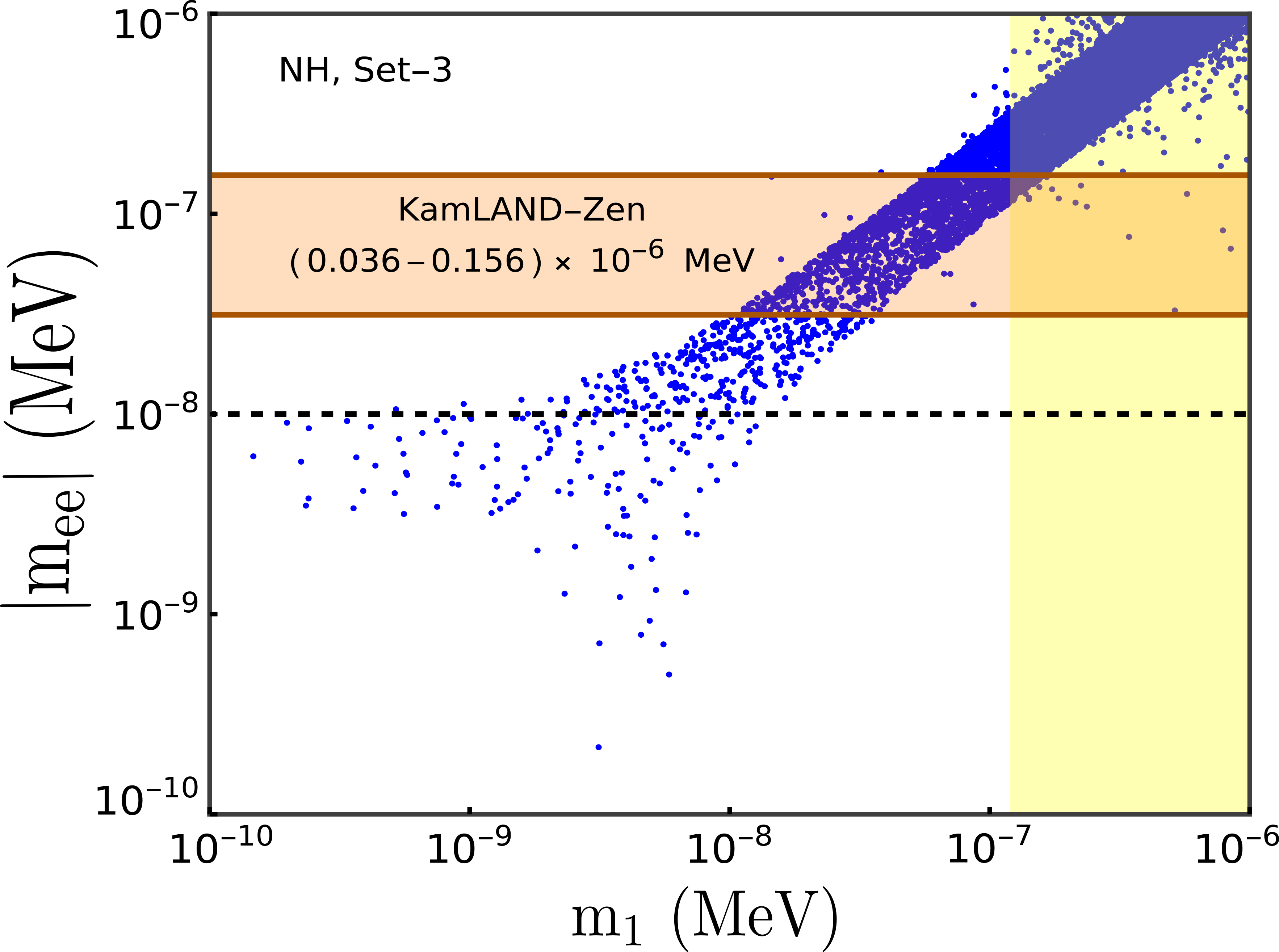}} \\
    \subcaptionbox{\label{fig:sub10}}{\includegraphics[width=5cm,height=5.2cm]{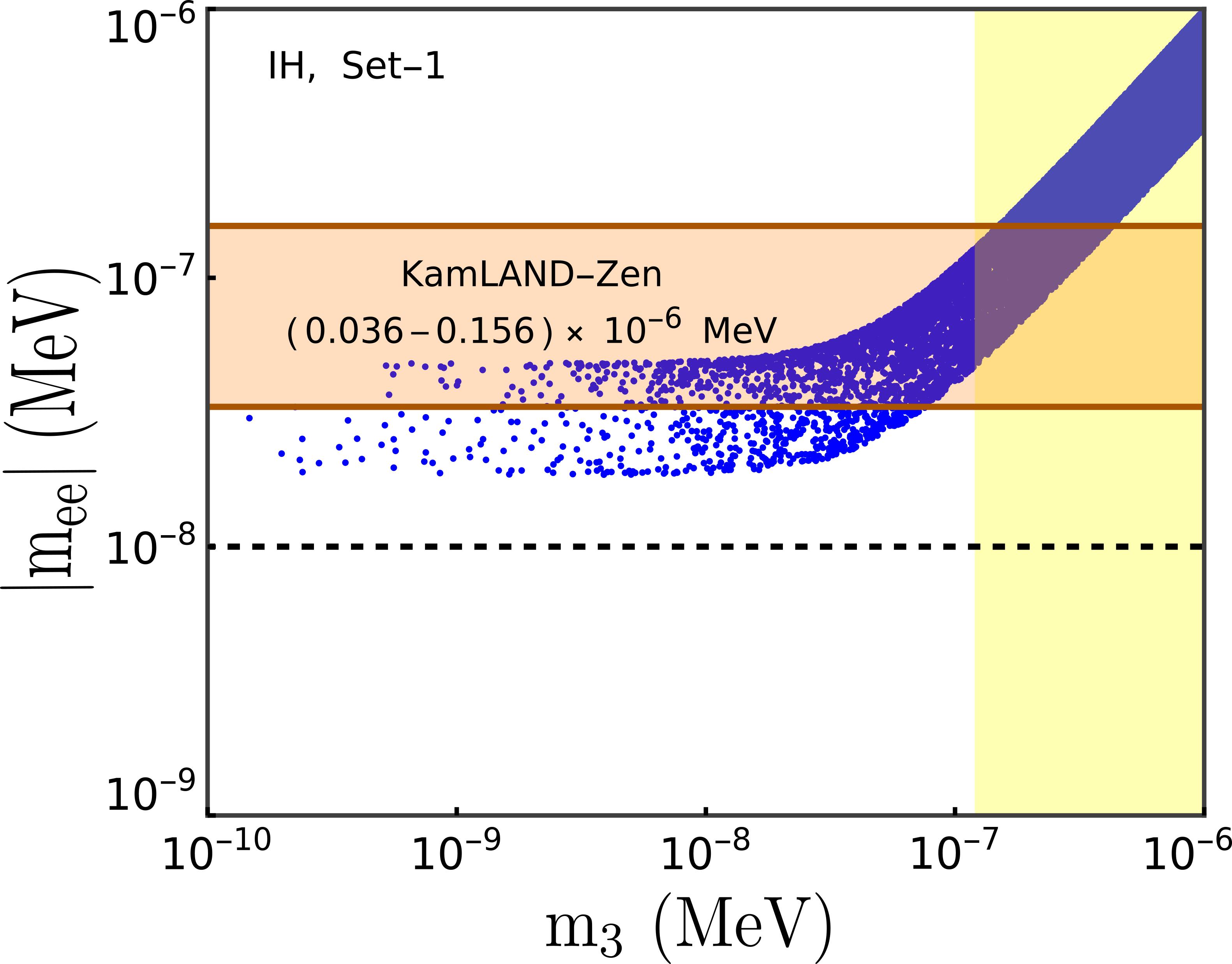}}
    \hspace{0.2cm}
    \subcaptionbox{\label{fig:sub11}}{\includegraphics[width=5cm,height=5.2cm]{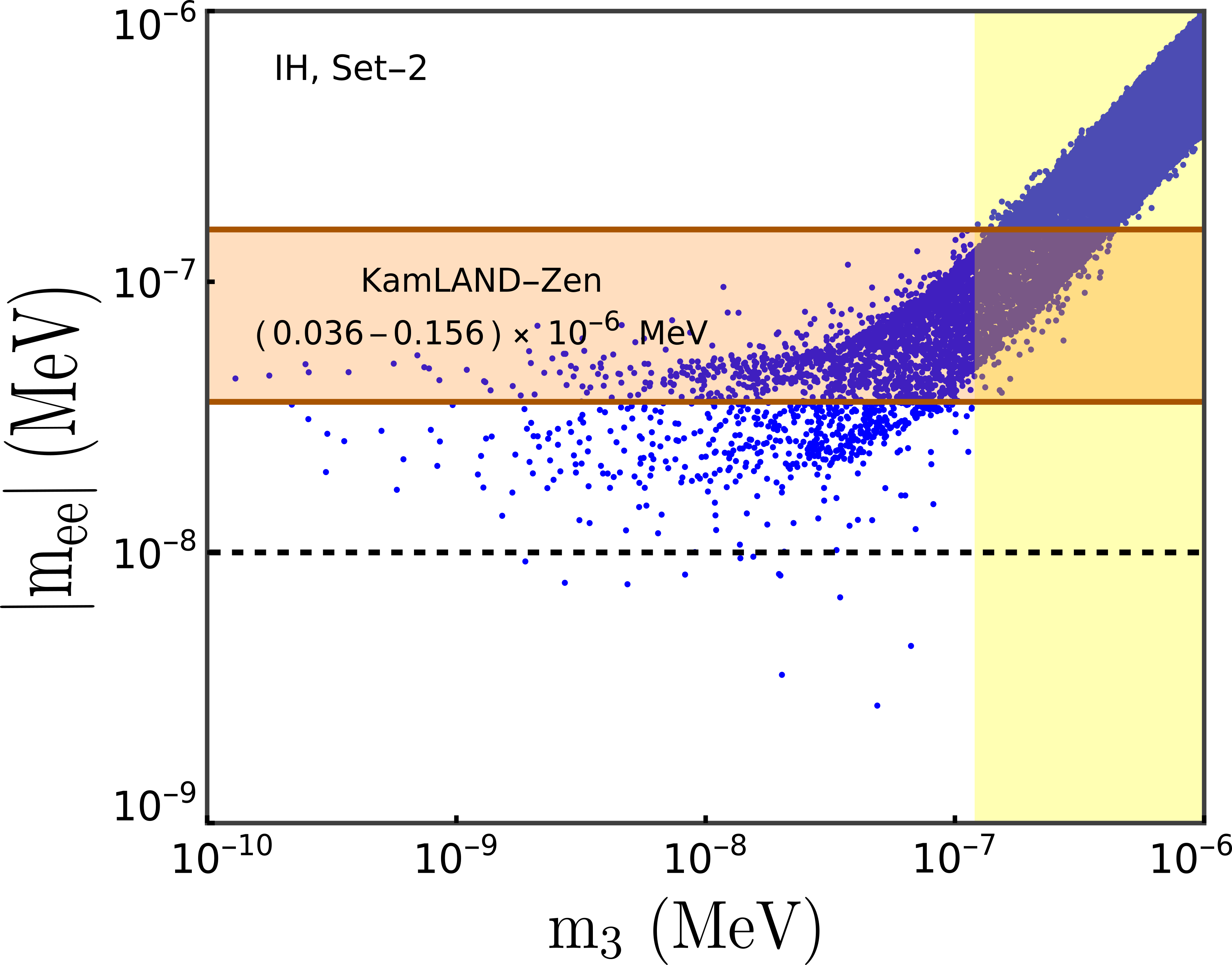}}
    \hspace{0.2cm}
    \subcaptionbox{\label{fig:sub12}}{\includegraphics[width=5cm,height=5.2cm]{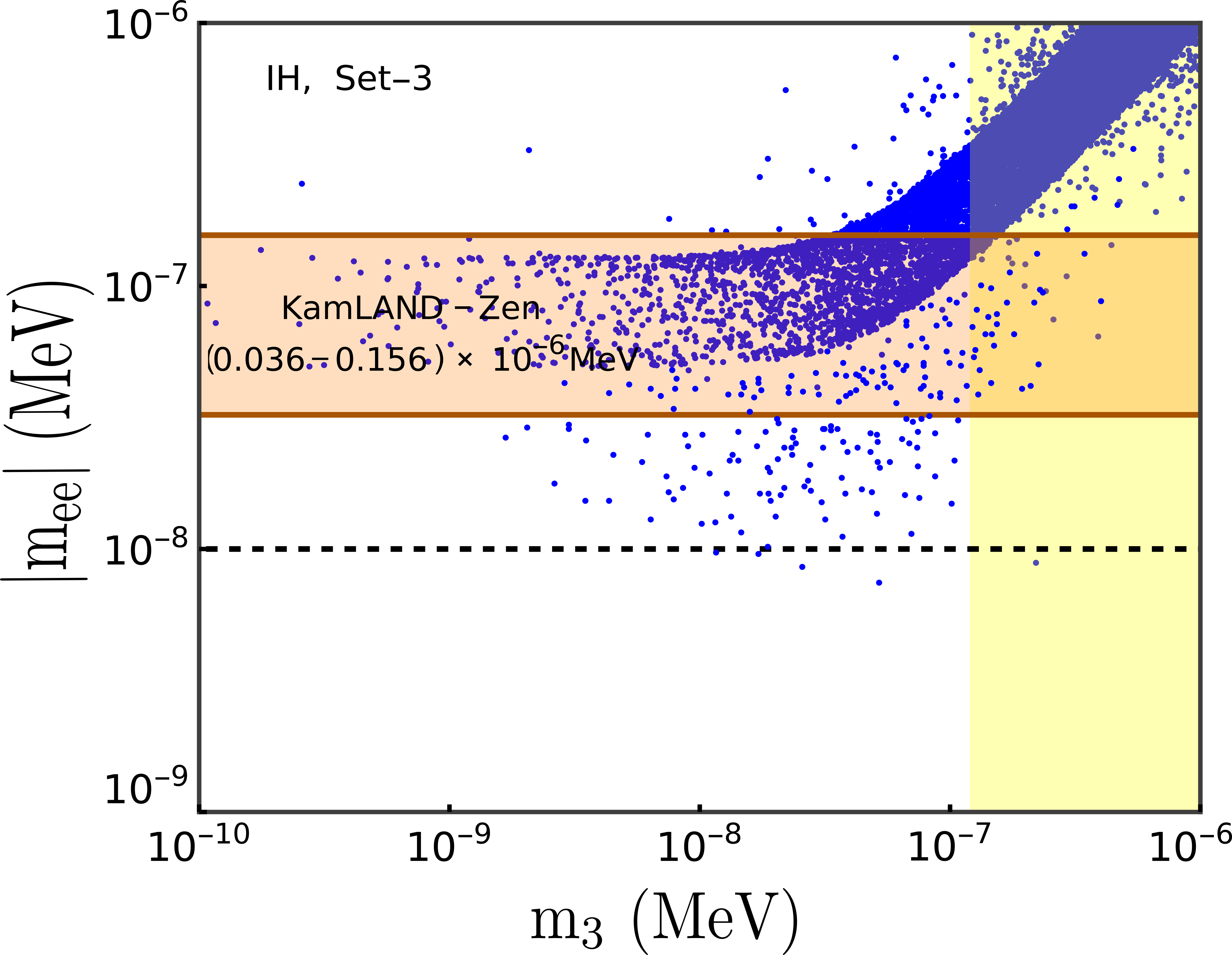}}
    \caption{Figure depicting the variation of effective neutrino mass $m_{ee}$ of $0\nu \beta \beta$ decay against the lightest neutrino mass state $m_1$ ($m_3$) for ${}^{136}\rm{Xe}$ isotope. The upper (lower) panel is for NH(IH) of light active neutrinos. The dotted black line represents the current experimental bound on $m_{ee}$ from KamLAND-Zen. The light yellow region corresponds to the cosmologically disfavored region where the sum of light neutrino mass states exceeds $0.12\times10^{-6}$ MeV. The $m_{ee}$ value in case of set-1 (the extreme left plots) include contribution from sterile neutrinos in the mass range $(100 - 10^8)$ MeV. For sets - 2 and 3, the $m_{ee}$ values include sterile neutrinos in the mass range $(1-10^6)$ MeV and $(10^{-4} - 10^{4})$ MeV, respectively (shown in middle and extreme right plots).}
    \label{fig-mee}
\end{figure}

Now using our model setup and substituting the appropriate amplitude expression from 
Eq. (\ref{mass-dependent-amp}), we have obtained the plots for $0\nu\beta\beta$ half-life for ${}^{136}\rm{Xe}$ isotope as shown in Fig. (\ref{fig-thalf}). Here, the upper (lower) panel is for NH (IH) of light active neutrinos.  The horizontal black line represents the current half-life sensitivity from the KamLand-Zen experiment. The yellow region indicates the cosmologically disfavored region where the sum of the light neutrino states exceeds the experimentally observed value, i.e., $\sum m_i < 0.12 \times 10^{-6}~ \rm{MeV}$. After calculating the $0\nu \beta \beta$ decay half-life using Eq. (\ref{half-life}), we have used that $\thalf$ value in Eq. (\ref{mee}) and calculated the effective neutrino mass $m_{ee}$, which is shown in Fig.(\ref{fig-mee}). The nature of the $m_{ee}$ is consistent with the theoretical prediction as per the relation $|m_{ee}|^2 \propto 1/\thalf$. In Fig. (\ref{fig-mee}), the orange band represents the current sensitivity on $m_{ee}$ from the KamLAND-Zen experiment \cite{KamLAND-Zen:2022tow}. The dotted black line represents the future sensitivity from experiments like LEGEND \cite{LEGEND:2017cdu}, nEXO \cite{nEXO:2017nam, nEXO:2018ylp}, and CUPID \cite{CUPID:2019imh, Armengaud:2019loe}.

Now we shall discuss the contribution of sterile neutrinos in detail, according to their various mass ranges. First, considering  set-1, where the sterile neutrinos lie in the mass range from $(100-10^8)$ MeV with very low mixing angles, the half-life value exceeds $10^{29}$ yr for very low mass of the lightest active neutrino i.e., $m_1 \sim (10^{-10} - 10^{-9})$ MeV for NH, followed by a pronounced upward spread and then a sharp decrease beyond $m_1 \sim  10^{-8}$ MeV towards the degenerate region (shown in Fig.(\ref{fig:sub1}). Similarly, for IH, the $\thalf$ remains almost constant, forming an extended band around $10^{27}$ yr for very low mass of $m_3$ up to $10^{-8}$ MeV and beyond this decreases sharply (shown in Fig.(\ref{fig:sub4})). The corresponding inverse behavior regarding $m_{ee}$ is depicted in figures \ref{fig:sub7} and \ref{fig:sub10} for NH and IH of light active neutrinos, respectively. For smaller values of $m_1$ ($\leq 10^{-9}$ MeV), $m_{ee}$ remains suppressed followed by a partial cancellation region between $(10^{-9} - 10^{-8})$ MeV that allows further lower value of it. Beyond $m_1$ ($\sim 10^{-8}$ MeV), $m_{ee}$ exhibit a steady growth. Similarly, for IH, $m_{ee}$ attains a lower value between $(1.87 - 4.6)\times 10^{-8}$ MeV, forming a well-defined band up to $m_3 \sim 10^{-8}$ MeV, and beyond this starts increasing smoothly towards the degenerate region of light active neutrinos. \textit{The well-defined coherent pattern observed in the case of set-1 resembles the SM light neutrino contribution, thus inferring about the fact that heavier sterile neutrinos might have integrated out and have a vanishing contribution in $\thalf$ and $m_{ee}$.} The result of $\thalf$ and $m_{ee}$ obtained considering the sterile neutrinos from set - 2 and 3, along with the SM neutrinos, also follow a similar fashion as that of set-1 but with a different scale. For NH, the maximum $\thalf$ value lies between $10^{28}-10^{29}$ yrs for both set 2 and 3, before the upward spread as shown in figures \ref{fig:sub2} and \ref{fig:sub3}. However, significant changes in the upward spread region can be noticed, more specifically when sterile neutrinos are considered in the mass range $(10^{-4} - 10^{4})$ MeV in set-3, the upward spread region is least populated, signifying the least cancellation among the parameter space. This can be clearly inferred from the least prominent cancellation region of $m_{ee}$ in the Fig.(\ref{fig:sub9}). For the IH case, as can be seen from figures \ref{fig:sub5} and \ref{fig:sub6}, the $\thalf$ values are not organized into a sharp band; they show a broader dispersion of points around the main trend. A similar kind of less confined distribution of $m_{ee}$ values is also observed in the case of set-2 and 3 (shown in figures \ref{fig:sub11} and \ref{fig:sub12}). The appearance of such scattered points may be attributed to the indication of enhanced contribution of sterile neutrinos from the mass range $(1-10^6)$ MeV and $(10^{-4}-10^4)$ MeV.

So far as the current sensitivity of $\thalf$ from the KamLAND-Zen experiment is concerned, some portion of our obtained result is found to be forbidden by the cosmological bound for all. For NH, the major portion of $\thalf$ is approaching the future sensitivity from experiments like LEGEND and its variants. This suggests that our parameter space in the NH condition is more relevant for next-generation experiments. For IH, the $\thalf$ values obtained considering sterile neutrinos from sets 1 and 2 are found to comply well with the KamLAND-Zen experimental bound, but some portion of the $\thalf$ value obtained including sterile neutrinos from set-3 is found to be ruled out by the current experimental limit. Similarly, the results regarding $m_{ee}$ for NH are found to comply with the current KamLAND-Zen bound as well as future sensitivities from experiments like CUPID, nEXO, and LEGEND. Specifically, the KamLAND-Zen bound overlaps with the $m_{ee}$ values in the quasi-degenerate region for light active neutrinos. For IH, the major portion of the $m_{ee}$ results obtained considering the sterile neutrinos from sets 1 and 2 are found to be constrained by the cosmological bound in the quasi-degenerate region of active neutrinos. As can be seen from Fig.(\ref{fig:sub10}), although the results satisfy the KamLAND-Zen bound, they may become disfavored as next-generation experiments reach their projected sensitivity. A similar scenario is followed by the $m_{ee}$ results from set 2, but in this case, very little parameter space remains available for future sensitivity. However, a completely different scenario is observed in the case of the contribution of sterile neutrinos from set 3. As can be seen in Fig. (\ref{fig:sub12}), the $m_{ee}$ value corresponding to the quasi-degenerate region of active neutrinos is completely disfavored by both the current KamLAND-Zen bound and cosmological bound. Although its rest portion are in good agreement with the KamLAND-Zen bound, it is still in a conflict zone from the future sensitivity perspective. 
 
Since we are focusing on keV sterile neutrinos, set 3 is more important for our analysis. Here, the mass of all three sterile neutrinos lies in the range $(10^{-4}-10^4)$ MeV, among which the realization of keV-scale sterile neutrino seems possible naturally. So, depending on the value of phases and mixing angles, various combinations of the sterile neutrino mass can be realized that lead to the desired value of half-life and effective neutrino mass. For illustrative purposes, we have shown some parameter sets in the table (\ref{tab:points}) for both NH and IH. Here, for this particular set of mixing angles and phase values, the different combinations of sterile neutrino masses give the $m_{ee}$ and $\thalf$ value that complies with the KamLAND-Zen bound.
\begin{table}[ht]
\centering
\renewcommand{\arraystretch}{1.3}
\resizebox{\textwidth}{!}{%
\begin{tabular}{|c|c|c|}
\hline
\multicolumn{3}{|c|}{\textbf{NH}} \\
\hline
\textbf{Phases} & \textbf{Mixing angles} & \textbf{Observables} \\
\hline
$\phi_{12}=0.4944,\ \phi_{13}=1.9239$ &
$\theta_{14}=1.62\times10^{-3},\ \theta_{24}=1.56\times10^{-2},\ \theta_{34}=2.66\times10^{-1}$ &
$M_1=3.58~\text{GeV},\ M_2=9.08~\text{keV}$ \\
$\phi_{14}=2.3836,\ \phi_{15}=1.4786$ &
$\theta_{15}=3.57\times10^{-4},\ \theta_{25}=9.80\times10^{-2},\ \theta_{35}=1.28\times10^{-1}$ &
$M_3=2.03~\text{MeV},\ |m_{ee}|=0.009~\text{eV}$ \\
$\phi_{16}=2.8915$ &
$\theta_{16}=7.99\times10^{-3},\ \theta_{26}=9.61\times10^{-3},\ \theta_{36}=5.56\times10^{-1}$ &
$\thalf=1.20\times10^{28} ~\rm{yr}$ \\
\hline
$\phi_{12}=4.4537,\ \phi_{13}=2.1516$ &
$\theta_{14}=1.39\times10^{-4},\ \theta_{24}=7.63\times10^{-3},\ \theta_{34}=4.16\times10^{-1}$ &
$M_1=8.39~\text{GeV},\ M_2=0.11~\text{GeV}$ \\
$\phi_{14}=0.6760,\ \phi_{15}=1.9352$ &
$\theta_{15}=7.63\times10^{-3},\ \theta_{25}=1.11\times10^{-2},\ \theta_{35}=1.69\times10^{-1}$ &
$M_3=14.31~\text{keV},\ |m_{ee}|=0.012~\text{eV}$ \\
$\phi_{16}=1.4643$ &
$\theta_{16}=1.41\times10^{-4},\ \theta_{26}=1.66\times10^{-2},\ \theta_{36}=8.19\times10^{-1}$ &
$\thalf=6.39\times10^{27} ~\rm{yr}$ \\
\hline
\multicolumn{3}{|c|}{\textbf{IH}} \\
\hline
$\phi_{12}=1.7805,\ \phi_{13}=0.8473$ &
$\theta_{14}=1.55\times10^{-3},\ \theta_{24}=6.47\times10^{-3},\ \theta_{34}=1.37\times10^{-1}$ &
$M_1=1.74~\text{GeV},\ M_2=13.68~\text{keV}$ \\
$\phi_{14}=2.8374,\ \phi_{15}=3.2356$ &
$\theta_{15}=8.98\times10^{-3},\ \theta_{25}=1.66\times10^{-2},\ \theta_{35}=9.87\times10^{-2}$ &
$M_3=126.2~\text{MeV},\ |m_{ee}|=0.038~\text{eV}$ \\
$\phi_{16}=1.4534$ &
$\theta_{16}=6.17\times10^{-4},\ \theta_{26}=1.05\times10^{-2},\ \theta_{36}=1.21\times10^{-1}$ &
$\thalf=7.25\times10^{26} ~\rm{yr}$ \\
\hline
$\phi_{12}=1.4217,\ \phi_{13}=4.29$ &
$\theta_{14}=5
87\times10^{-3},\ \theta_{24}=1.52\times10^{-3},\ \theta_{34}=3.13\times10^{-1}$ &
$M_1=8.54~\text{GeV},\ M_2=287.5~\text{MeV}$ \\
$\phi_{14}=1.7169,\ \phi_{15}=5.0456$ &
$\theta_{15}=4.12\times10^{-3},\ \theta_{25}=8.46\times10^{-2},\ \theta_{35}=3.58\times10^{-1}$ &
$M_3=12.54~\text{keV},\ |m_{ee}|=0.044~\text{eV}$ \\
$\phi_{16}=4.0802$ &
$\theta_{16}=9.02\times10^{-3},\ \theta_{26}=7.71\times10^{-2},\ \theta_{36}=8.35\times10^{-1}$ &
$\thalf=5.48\times10^{26} ~\rm{yr}$ \\
\hline
\end{tabular}
}
\caption{Table showing the values of sterile neutrino masses and corresponding half-life and effective neutrino mass obtained for a certain set of mixing angle values and CP violating phases. These points correspond to the scenario involving the sterile neutrinos from set-3.}
\label{tab:points}
\end{table}
\section{Conclusion}\label{4-con}
In this work, we investigate the contribution of sterile neutrinos in the keV mass range to
the $0\nu\beta\beta$ decay using the $\chi$EFT approach. For this, we have extended the SM by adding three sterile neutrinos, which are included in the lepton mixing sector through a $6\times6$ mixing matrix $\mathcal{U}$. This mixing matrix $\mathcal{U}$ is parametrized in terms of mixing angles $\theta_{ij}$ and CP-violating phases $\phi_{ij}$. Then, using the exact seesaw relation, we have obtained the analytical expressions for the added sterile neutrino mass states $M_i$ in terms of the parametric expression of $\mathcal{U}$. Based on those analytical expressions, our model parameter space comprises three active-active mixing angles (having well-defined values from the oscillation experiments), nine active-sterile mixing angles, five CP-violating phases, and the lightest mass state of SM active neutrinos $m_{\rm lightest}$. We have taken three different sets for the active-sterile mixing angles and an independent variation of phases from $0$ to $2\pi$. The analytical expressions upon varying with $m_{\rm lightest}$ yield different mass ranges corresponding to different sets of mixing angles. We got all the $M_i$ lie in the range $(100 - 10^8)$ MeV, $(1 - 10^6)$ MeV and $(10^{-4} - 10^{4})$ MeV in case of set-1,2 and 3 respectively. 

Now, based on their different mass range, they are included in the calculation of $\thalf$ and $m_{ee}$ of $0\nu\beta\beta$ decay in the limit of the $\chi$EFT approach. The effective formula for the decay amplitude considered in the $\chi$EFT framework includes all the contributions thoroughly from both active neutrinos and sterile neutrinos from different momentum regions defined according to their mass scale and accumulate appropriate combination of NMEs for the calculation. In our analysis, we have taken the NME values computed from the nuclear shell model for the isotope ${}^{136}\rm Xe$. Then, using the working formula for $\thalf$ and $m_{ee}$, we have studied the combined contribution of active neutrinos and sterile neutrinos. The obtained results are then plotted against the lightest active neutrino mass state $m_1(m_3)$ for NH(IH). When the sterile neutrinos from set-1 are considered along with the active neutrinos, the obtained plots for $\thalf$ and $m_{ee}$ resemble more the standard three neutrino contribution, which means we are getting dominant contribution from the active neutrinos, which is clearly visible from the prominent cancellation region in case of NH and extended band in case of IH. However, the prominence of the cancellation is found to decrease when the contribution of sterile neutrinos from sets 2 and 3 is considered. Specifically, the least cancellation effect from set-3 sterile neutrinos indicates their enhanced contribution in $m_{ee}$. Similarly, in the case of IH, the contribution of sterile neutrinos from sets 2 and 3 makes the band distorted with the appearance of scattered points around the main trend.

From the keV scale dark matter constituent point of view, set-3 is of primary interest as all the added sterile neutrino masses lie in the range $(10^{-4} - 10^4)$ MeV here. So this is the most probable region where keV scale sterile neutrino can come into the scenario. However, depending upon the phase values and mixing angles, $M_i$ can contribute to $\thalf$ and $m_{ee}$ in different mass scales. This fact can be noticed from the table (\ref{tab:points}), where we have illustrated a certain set of mixing angles and phase values and corresponding sterile neutrino masses. There is an appearance of at least one sterile neutrino from the keV scale, and the $\thalf$ and $m_{ee}$ values obtained in this regard are also found to satisfy the KamLAND-Zen bound.

From experimentally testable point of view, low scale variant of type-I seesaw is more appealing. There are different theoretical frameworks and experimental searches according to their mass scale. For example, eV-scale sterile neutrinos can be probed in short-baseline experiments \cite{Conrad:2012qt, Danilov:2022str, Katori:2014vka}. keV scale neutrinos contextualised in $\nu$MSM and are getting probed from cosmological data \cite{Asaka:2005an, Boyarsky:2009ix}. Sterile neutrinos within mass range MeV - TeV are getting more attention in collider searches \cite{Abada:2022wvh}, in rare meson decays \cite{Atre:2009rg}, and also through precision measurement of electroweak observables \cite{Antusch:2015mia}. Popular models like $A_4$ symmetry \cite{Lindner:2010wr,CarcamoHernandez:2019kjy} or softly broken $L_e-L_\mu-L_\tau$ symmetry \cite{Barry:2011wb} also make it possible to realize the low-scale variant of the type-I seesaw. In view of these existing realizations of low-scale type-I seesaw, the parameter space explored in our model, where the added sterile neutrinos attain a mass from $10^{-4}$ MeV to $10^8$ MeV, can offer a promising platform for the realization of low scale variant of type-I seesaw. 
  
\appendix
\section{Structure of amplitude in different momentum region}\label{amp-formula}
\subsubsection{Region 1: $2 \text{ GeV} \le M_i$ }\label{heavy}
In this region, the massive neutrinos can be integrated out at the quark level, leading to a local LNV dimension-nine operator (containing 4 quarks and 2 electrons) that scales as $M_i^{-2}$. The corresponding amplitude is given by,
\begin{align}
\amp^{(9)}(M_i) = -2\eta(\mu_0,M_i) \frac{m_\pi^2}{M_i^2}\Bigg[&\frac{5}{6}g_1^{\pi\pi}\left(\mathcal{M}_{GT,sd}^{PP}+\mathcal{M}_{T,sd}^{PP}\right) \nonumber \\
+&\frac{g_1^{\pi N}}{2}\left(\mathcal{M}_{GT,sd}^{AP}+\mathcal{M}_{T,sd}^{AP}\right)
-\frac{2}{g_A^2}g_1^{NN}\mathcal{M}_{F,sd}\Bigg]\,,
\label{amp-dim9}
\end{align}
where the QCD evolution at the renormalization scale $\mu_0\simeq 2$ GeV in terms of bottom ($m_{\text{bottom}}$) and top ($m_{\text{top}}$) quark masses is given by,
\begin{align*}
\eta(\mu_0,M_i) =\left\{
  \begin{array}{@{}ll@{}}
   \left(\frac{\alpha_s(M_i)}{\alpha_s(\mu_0)}\right)^{6/25} & M_i\leq m_{\text{bottom}} \\
\left(\frac{\alpha_s(m_{\text{bottom}})}{\alpha_s(\mu_0)}\right)^{6/25}\left(\frac{\alpha_s(M_i)}{\alpha_s(m_{\text{bottom}})}\right)^{6/23} & m_{\text{bottom}} \leq M_i\leq m_{\text{top}}\\
\left(\frac{\alpha_s(m_{\text{bottom}})}{\alpha_s(\mu_0)}\right)^{6/25}\left(\frac{\alpha_s(m_{\text{top}})}{\alpha_s(m_{\text{bottom}})}\right)^{6/23} \left(\frac{\alpha_s(M_i)}{\alpha_s(m_{\text{top}})}\right)^{2/7} & M_i \geq m_{\text{top}}
  \end{array}\right. \,,
\label{QCDparameter}
\end{align*}
  and  $\alpha_s(\mu)=\frac{2\pi}{\beta_0\log(\mu/\Lambda^{(n_f)})}$, with $\beta_0 = 11-\frac{2}{3}n_f$, is the strong coupling constant at one loop, and $\alpha_s(m_Z) = 0.1179$ gives $\Lambda^{(4,5,6)} \simeq \{119,\, 87,\, 43 \}$ MeV. $g_1^{\pi\pi}$, $g_1^{N\pi}$, and $g_1^{NN}$ denote the LECs, evaluated at the scale $\mu_0=2$ GeV, corresponding to the $\pi\pi$, $\pi N$, and $NN$ interactions, which are expected to be $\mathcal{O}(1)$. Among these three LECs, the values $g_1^{N\pi}$, and $g_1^{NN}$ are not sufficiently explored. However, different values of $g_1^{\pi\pi}$ have been estimated through the pionic coupling using LQCD calculations. To achieve a reliable estimate, we will consider the contribution of $g_1^{\pi\pi}$ and $g_1^{NN}$. Inspired by a factorization estimation, the values are given by $g_1^{NN}= (1+3 g_A^2)/4 $ and $g^{\pi\pi}_1=0.36$ \cite{Nicholson:2018mwc}. The values of NMEs are given in table~\ref{tab:NME}. 
\begin{table}
	\center
		\renewcommand{\arraystretch}{1.2}    
	\begin{tabular}[b]{|c|ccccc|}    
		\hline		
		&  $\mathcal M_{F,sd}$ & $\mathcal M_{GT,sd}^{AP}$ &$\mathcal M_{GT,sd}^{PP} $&$\mathcal M_{T,sd}^{AP} $&$\mathcal M_{T,sd}^{PP}$\\\hline
		$^{136}$Xe&-1.94 &-1.99 &0.74 &0.05 &-0.02\\\hline
	\end{tabular}
	\caption{Values of NMEs for ${}^{136}$Xe calculated by Shell-model.}
	\label{tab:NME}
\end{table}

\subsubsection{Region 2: $100 ~\rm{ MeV} \le M_i <  2 ~\rm{ GeV}$ }\label{intermediate}
In this mass range, the sterile neutrinos are to be kept as explicit degrees of freedom in $\chi$EFT. Its effects can be captured by including the mass dependence of the NMEs and LEC. The corresponding amplitude can be written
\begin{equation}
    \amp(M_i)=\amp^{\rm (ld)}(M_i)+\amp^{\rm (sd)}(M_i) =  -\mathcal{M}(M_i)-2 g_\nu^{NN}(M_i) m_\pi^2\frac{\mathcal{M}_{F,sd}}{g_A^2}\,,
    \label{amp-intermediate}
\end{equation}
where 
\begin{equation}
    \mathcal{M}(M_i)=\frac{\mathcal{M}_F(M_i)}{g_A^2}-\mathcal{M}_{GT}(M_i)-\mathcal{M}_T(M_i).
\end{equation}
In the limit of EFT, $\mathcal{M}(M_i)$ should get a linear dependence on $M_i$ for small mass, but they scale as $M_i^{-2}$ for heavy mass. The description in terms of NMEs no longer applies for masses $M_i \geq \mu_0 = 2$ GeV, as we integrate out the heavy neutrinos at the quark level in this case. In practice, it is useful to have an interpolation formula that describes the shell-model results for $M_i\leq 2$ GeV. The proposed interpolation formula has the following functional form
\begin{equation}
\mathcal{M}_{\mathrm{int}}(M_i) =\mathcal{M}(0)\frac{1}{1+M_i/m_a+(M_i/m_b)^2}\,,
\label{eq:interpolation}
\end{equation}
and the mass dependence of the LEC $g_\nu^{NN}(M_i)$ from the short-distance part, has the following functional form,
\begin{equation}
g_{\nu}^{NN}(M_i) = g_{\nu}^{NN}(0) \frac{1+ (M_i/m_c)^2\,{\rm sign}(m_d^2)}{1 + (M_i/m_c)^2(M_i/|m_d|)^2}\,,
\label{eq:gnu_int}
\end{equation}
where sign$(m_d^2) = m_d^2/|m_d^2|$, $g_{\nu}^{NN}(0) = -1.01\,\mathrm{fm}^2$ \cite{Jokiniemi:2021qqv}, and we set $m_c = 1$ GeV. The values of $m_d$ for $^{136}$Xe, as well as $m_a$, $m_b$, $\mathcal{M}(0)$ from the long-distance part, are given in table-(\ref{tab:NME1}).

\begin{table}
	\center
		\renewcommand{\arraystretch}{1.2}    
	\begin{tabular}[b]{|c|cccc|}    
		\hline		
		&   $m_a$ & $m_b$ & $m_d$ & $\mathcal{M}(0)$\\\hline
		$^{136}$Xe&157&221&146&2.7\\\hline
	\end{tabular}
	\caption{for ${}^{136}$Xe. $m_a,\,m_b,\,m_c$ are dimensionful, and are given in MeV here.} 
    \label{tab:NME1}
\end{table}
\subsubsection{Region 3: $ M_i <100 ~\rm{ MeV} $ }\label{light}
In this region, the sterile neutrinos have contributions from the hard and potential momentum region, similar to the SM neutrinos but with a relatively different importance. An additional significant contribution comes from the ultrasoft region. Instead of the naive long-distance part $\amp^{(\rm ld)}(M_i)$, a modified one, i.e., $\amp^{(\rm ld,<)}(M_i)$, is considered to avoid the double counting of a specific piece of neutrino-exchange amplitude. Although this low-momentum part of the neutrino propagator is already accounted for by the ultrasoft neutrinos, it also emerges inside the unexpanded long-range potential. A detailed explanation is provided in the Ref.\cite{Dekens:2024hlz}. The modified long-distance part of the amplitude avoids the overlap by removing the linear $M_i$ term and is given by,
\begin{equation}
    \amp^{(\rm ld,<)}(M_i) = -\left(\mathcal{M}(M_i) - M_i \left[\frac{d}{d M_i} \mathcal{M}(M_i)\right]_{M_i=0}\right)\,,
    \label{amp1-light}
\end{equation}
where we use the functional form given in Eq.~\eqref{eq:interpolation} for $\mathcal{M}$. 

Now the ultrasoft contribution of amplitude is given by,
\begin{equation}\label{eq:amp-usoft}
\begin{aligned}
\amp^{\rm (usoft)}(M_i) &=& 
2\frac{R_A}{\pi g_A^2}   \sum_{n}   \langle 0^+_f|\mathcal J^\mu |1^+_n\rangle \langle 1^+_n| \mathcal J_\mu |0^+_i\rangle  \Big(f(M_i, \Delta E_1) +f(M_i,\Delta E_2)\Big)  \,,
\end{aligned}
\end{equation}
with
\begin{eqnarray}\label{eq:amp-usoft1}
f(m,E) = \begin{cases}
-2\left[ E\left(1+ \log \frac{\mu_{us}}{m} \right)  +\sqrt{m^2 - E^2} \left(\frac{\pi}{2}-\tan^{-1}\, \frac{ E}{\sqrt{m^2 -E^2}}\right) \right]\,,
&{\rm if}\,  m > E\,, \\
-2\left[ E \left(1+\log \frac{\mu_{us}}{m} \right)  -\sqrt{ E^2-m^2} \log \frac{ E+\sqrt{ E^2-m^2}}{m}  \right]\,,& {\rm if}\,  m \le E\,,
\end{cases}
\end{eqnarray}
where $\Delta E_{1,2} = E_{1,2}+E_n-E_i$, and $E_{i}$, $E_{n}$, $E_f$ are the energies of the initial, intermediate, and final state, respectively. $E_{1,2}$ stands for the electron energies, and we set the renormalization scale $\mu_{us}=m_\pi$.
To achieve a dependence on the initial and final states, the sum over the complete set of intermediate states can be approximated as follows:
\begin{align}
\sum_{n} \langle 0_f^+|\mathcal{J}^\mu|1_n^+\rangle\langle 1_n^+|\mathcal{J}_\mu|0_i^+\rangle
&\;\simeq\; \frac{1}{4}\Big[
   g_V^2\,\langle 0_f^+|\tau^+|1_n^+\rangle\langle 1_n^+|\tau^+|0_i^+\rangle  \nonumber \\
&\qquad\qquad\qquad
   -\,g_A^2\,\langle 0_f^+|\tau^+\boldsymbol{\sigma}|1_n^+\rangle\cdot\langle 1_n^+|\tau^+\boldsymbol{\sigma}|0_i^+\rangle
\Big]\nonumber\\
&\;\simeq\; -\frac{g_A^2}{4}  \langle 0 ^+_f |  \tau^+ \boldsymbol{\sigma}  | 1^+_n\rangle \cdot  \langle 1^+_n |  \tau^+  \boldsymbol{\sigma} | 0^+_i \rangle\,,
\label{eq:amp-usoft2}
\end{align}
where $\langle 0_f^+|\tau^+|1_n^+\rangle$ is the Fermi transition matrix elements and $\langle 1^+_n |  \tau^+  \boldsymbol{\sigma} | 0^+_i \rangle$ is the Gamow–Teller (GT) transition matrix element. The Fermi transitions can be neglected as they are highly suppressed by isospin conservation apart from small isospin-breaking corrections.
By knowing the intermediate state energies $E_n$ and the value of the nuclear matrix elements $\langle 0^+_f | \mathcal J_\mu | 1^+_n\rangle$, we can calculate $\amp^{\rm (usoft)}(M_i)$. The matrix elements involving intermediate states for $^{136}$Xe are given in the table-(\ref{tab:usoftNMEforXe}).
\begin{table*}[htbp!]
\setlength{\tabcolsep}{2pt}
\renewcommand{\arraystretch}{1.3}    \centering
\footnotesize
    \begin{tabular}{|c|c|c|}
    \hline
    $\frac{E_n-E_i}{\rm MeV}$&  $\langle 1^+_n| \boldsymbol{\sigma}\tau^+ | 0^+_i\rangle$& $\langle 0^+_f| \boldsymbol{\sigma}\tau^+ | 1^+_n\rangle$\\\hline
 0.17 & 1. & 0.13 \\
 0.63 & -0.19 & -0.0063 \\
 0.89 & -0.25 & -0.016 \\
 1.02 & 0.3 & 0.036 \\
 1.05 & 0.23 & 0.025 \\
 1.1 & -0.13 & -0.00076 \\
 1.2 & 0.12 & -0.0052 \\
 1.3 & 0.16 & -0.0028 \\
 1.4 & -0.23 & -0.0098 \\
 1.5 & 0.2 & -0.012 \\
 1.6 & -0.36 & 0.0084 \\
 1.7 & -0.24 & 0.00058 \\
 1.9 & 0.22 & 0.011 \\
 2.0 & 0.34 & 0.007 \\
 2.2 & 0.35 & 0.006 \\
 2.3 & -0.49 & -0.0086 \\
 2.6 & 0.62 & 0.021 \\
 2.7 & -0.91 & -0.024 \\
 2.9 & 0.37 & 0.0064 \\
 3.1 & 0.3 & 0.0013
  \\\hline
    \end{tabular}
\hfill    
        \begin{tabular}{|c|c|c|}
    \hline
    $\frac{E_n-E_i}{\rm MeV}$&  $\langle 1^+_n| \boldsymbol{\sigma}\tau^+ | 0^+_i\rangle$& $\langle 0^+_f| \boldsymbol{\sigma}\tau^+ | 1^+_n\rangle$\\\hline
 3.3 & 0.39 & -0.0013 \\
 3.6 & 0.39 & 0.0021 \\
 3.8 & 0.45 & -0.013 \\
 4.0 & -0.44 & -0.0032 \\
 4.3 & -0.35 & -0.0038 \\
 4.6 & -0.36 & -0.0067 \\
 4.8 & 0.44 & 0.0083 \\
 5.1 & 0.44 & 0.0066 \\
 5.4 & -0.55 & -0.0093 \\
 5.7 & 0.63 & 0.012 \\
 6.1 & 0.85 & 0.013 \\
 6.3 & -1.2 & -0.016 \\
 6.7 & -1.3 & -0.014 \\
 7.0 & -1.9 & -0.016 \\
 7.3 & 3.1 & 0.023 \\
 7.5 & -4. & -0.028 \\
 7.7 & 2.6 & 0.017 \\
 8.1 & 1.4 & 0.0091 \\
 8.4 & -1. & -0.0057 \\
 8.8 & -0.93 & -0.0064 
  \\\hline
    \end{tabular}
\hfill
        \begin{tabular}{|c|c|c|}
    \hline
 $\frac{E_n-E_i}{\rm MeV}$&  $\langle 1^+_n| \boldsymbol{\sigma}\tau^+ | 0^+_i\rangle$& $\langle 0^+_f| \boldsymbol{\sigma}\tau^+ | 1^+_n\rangle$\\\hline
 9.1 & 0.8 & 0.0038 \\
 9.4 & 0.59 & 0.0014 \\
 9.8 & -0.5 & 0.0027 \\
 10.1 & 0.35 & -0.0027 \\
 10.5 & 0.26 & -0.00053 \\
 10.9 & -0.22 & -0.00021 \\
 11.3 & 0.17 & -0.00037 \\
 11.7 & -0.16 & -0.00054 \\
 12.0 & -0.16 & -0.001 \\
 12.4 & 0.14 & 0.00092 \\
 12.8 & 0.12 & -0.00014 \\
 13.1 & 0.092 & -0.0004 \\
 13.5 & -0.079 & -0.00019 \\
 13.9 & 0.071 & -0.00026 \\
 14.2 & -0.07 & 0.000031 \\
 14.6 & -0.035 & 0.00021 \\
 15.1 & -0.051 & -0.00015 \\
 16.2 & -0.039 & 0.00011 \\
 17.3 & -0.043 & -0.000091 \\
 17.7 & 0.11 & -0.000029 
  \\\hline
    \end{tabular}    
    \caption{Values of the first-order nuclear matrix elements in Eq.\ \eqref{eq:amp-usoft2}, that enter the $0\nu\beta\beta$ of $^{136}$Xe.}
    \label{tab:usoftNMEforXe}
\end{table*}

For short-distance contribution, we can use our previously defined formula: 
\begin{equation}
\amp^{\rm (sd)}(M_i) = -2 g_\nu^{NN}(M_i) m_\pi^2\frac{\mathcal{M}_{F,sd}}{g_A^2}.
\label{amp3-sd}
\end{equation}
where only the hadronic matrix element $g_\nu^{NN}(M_i)$ is $M_i$ dependent and can be evaluated using Eq.(\ref{eq:gnu_int}).
\bibliographystyle{unsrt}
\bibliography{ref-3}

@article{Bulbul:2014sua,
    author = "Bulbul, Esra and Markevitch, Maxim and Foster, Adam and Smith, Randall K. and Loewenstein, Michael and Randall, Scott W.",
    title = "{Detection of An Unidentified Emission Line in the Stacked X-ray spectrum of Galaxy Clusters}",
    eprint = "1402.2301",
    archivePrefix = "arXiv",
    primaryClass = "astro-ph.CO",
    doi = "10.1088/0004-637X/789/1/13",
    journal = "Astrophys. J.",
    volume = "789",
    pages = "13",
    year = "2014"
}

@article{Ruchayskiy:2015onc,
    author = "Ruchayskiy, Oleg and Boyarsky, Alexey and Iakubovskyi, Dmytro and Bulbul, Esra and Eckert, Dominique and Franse, Jeroen and Malyshev, Denys and Markevitch, Maxim and Neronov, Andrii",
    title = "{Searching for decaying dark matter in deep XMM{\textendash}Newton observation of the Draco dwarf spheroidal}",
    eprint = "1512.07217",
    archivePrefix = "arXiv",
    primaryClass = "astro-ph.HE",
    doi = "10.1093/mnras/stw1026",
    journal = "Mon. Not. Roy. Astron. Soc.",
    volume = "460",
    number = "2",
    pages = "1390--1398",
    year = "2016"
}

@article{Boyarsky:2014ska,
    author = "Boyarsky, Alexey and Franse, Jeroen and Iakubovskyi, Dmytro and Ruchayskiy, Oleg",
    title = "{Checking the Dark Matter Origin of a 3.53 keV Line with the Milky Way Center}",
    eprint = "1408.2503",
    archivePrefix = "arXiv",
    primaryClass = "astro-ph.CO",
    doi = "10.1103/PhysRevLett.115.161301",
    journal = "Phys. Rev. Lett.",
    volume = "115",
    pages = "161301",
    year = "2015"
}

@article{Boyarsky:2014jta,
    author = "Boyarsky, Alexey and Ruchayskiy, Oleg and Iakubovskyi, Dmytro and Franse, Jeroen",
    title = "{Unidentified Line in X-Ray Spectra of the Andromeda Galaxy and Perseus Galaxy Cluster}",
    eprint = "1402.4119",
    archivePrefix = "arXiv",
    primaryClass = "astro-ph.CO",
    doi = "10.1103/PhysRevLett.113.251301",
    journal = "Phys. Rev. Lett.",
    volume = "113",
    pages = "251301",
    year = "2014"
}

@article{Iakubovskyi:2015dna,
    author = "Iakubovskyi, Dmytro and Bulbul, Esra and Foster, Adam R. and Savchenko, Denys and Sadova, Valentyna",
    title = "{Testing the origin of {\textasciitilde}3.55 keV line in individual galaxy clusters observed with XMM-Newton}",
    eprint = "1508.05186",
    archivePrefix = "arXiv",
    primaryClass = "astro-ph.HE",
    month = "8",
    year = "2015"
}

@article{Dodelson:1993je,
    author = "Dodelson, Scott and Widrow, Lawrence M.",
    title = "{Sterile-neutrinos as dark matter}",
    eprint = "hep-ph/9303287",
    archivePrefix = "arXiv",
    reportNumber = "FERMILAB-PUB-93-057-A",
    doi = "10.1103/PhysRevLett.72.17",
    journal = "Phys. Rev. Lett.",
    volume = "72",
    pages = "17--20",
    year = "1994"
}

@article{Bezrukov:2005mx,
    author = "Bezrukov, F. L.",
    title = "{nu MSM-predictions for neutrinoless double beta decay}",
    eprint = "hep-ph/0505247",
    archivePrefix = "arXiv",
    doi = "10.1103/PhysRevD.72.071303",
    journal = "Phys. Rev. D",
    volume = "72",
    pages = "071303",
    year = "2005"
}

@article{Agrawal:2021dbo,
    author = "Agrawal, Prateek and others",
    title = "{Feebly-interacting particles: FIPs 2020 workshop report}",
    eprint = "2102.12143",
    archivePrefix = "arXiv",
    primaryClass = "hep-ph",
    doi = "10.1140/epjc/s10052-021-09703-7",
    journal = "Eur. Phys. J. C",
    volume = "81",
    number = "11",
    pages = "1015",
    year = "2021"
}

@article{Drewes:2013gca,
    author = "Drewes, Marco",
    title = "{The Phenomenology of Right Handed Neutrinos}",
    eprint = "1303.6912",
    archivePrefix = "arXiv",
    primaryClass = "hep-ph",
    reportNumber = "TUM-HEP-881-13",
    doi = "10.1142/S0218301313300191",
    journal = "Int. J. Mod. Phys. E",
    volume = "22",
    pages = "1330019",
    year = "2013"
}

@article{Shaposhnikov:2006nn,
    author = "Shaposhnikov, Mikhail",
    title = "{A Possible symmetry of the nuMSM}",
    eprint = "hep-ph/0605047",
    archivePrefix = "arXiv",
    reportNumber = "CERN-PH-TH-2006-079",
    doi = "10.1016/j.nuclphysb.2006.11.003",
    journal = "Nucl. Phys. B",
    volume = "763",
    pages = "49--59",
    year = "2007"
}

@article{Kersten:2007vk,
    author = {Kersten, J{\"o}rn and Smirnov, Alexei Yu.},
    title = "{Right-Handed Neutrinos at CERN LHC and the Mechanism of Neutrino Mass Generation}",
    eprint = "0705.3221",
    archivePrefix = "arXiv",
    primaryClass = "hep-ph",
    doi = "10.1103/PhysRevD.76.073005",
    journal = "Phys. Rev. D",
    volume = "76",
    pages = "073005",
    year = "2007"
}

@article{KamLAND-Zen:2022tow,
    author = "Abe, S. and others",
    collaboration = "KamLAND-Zen",
    title = "{Search for the Majorana Nature of Neutrinos in the Inverted Mass Ordering Region with KamLAND-Zen}",
    eprint = "2203.02139",
    archivePrefix = "arXiv",
    primaryClass = "hep-ex",
    doi = "10.1103/PhysRevLett.130.051801",
    journal = "Phys. Rev. Lett.",
    volume = "130",
    number = "5",
    pages = "051801",
    year = "2023"
}

@article{GERDA:2020xhi,
    author = "Agostini, M. and others",
    collaboration = "GERDA",
    title = "{Final Results of GERDA on the Search for Neutrinoless Double-$\beta$ Decay}",
    eprint = "2009.06079",
    archivePrefix = "arXiv",
    primaryClass = "nucl-ex",
    doi = "10.1103/PhysRevLett.125.252502",
    journal = "Phys. Rev. Lett.",
    volume = "125",
    number = "25",
    pages = "252502",
    year = "2020"
}

@article{Minkowski:1977sc,
    author = "Minkowski, Peter",
    title = "{$\mu \to e\gamma$ at a Rate of One Out of $10^{9}$ Muon Decays?}",
    reportNumber = "Print-77-0182 (BERN)",
    doi = "10.1016/0370-2693(77)90435-X",
    journal = "Phys. Lett. B",
    volume = "67",
    pages = "421--428",
    year = "1977"
}

@article{Mohapatra:1979ia,
    author = "Mohapatra, Rabindra N. and Senjanovic, Goran",
    title = "{Neutrino Mass and Spontaneous Parity Nonconservation}",
    reportNumber = "MDDP-TR-80-060, MDDP-PP-80-105, CCNY-HEP-79-10",
    doi = "10.1103/PhysRevLett.44.912",
    journal = "Phys. Rev. Lett.",
    volume = "44",
    pages = "912",
    year = "1980"
}

@article{Yanagida:1980xy,
    author = "Yanagida, Tsutomu",
    title = "{Horizontal Symmetry and Masses of Neutrinos}",
    reportNumber = "TU-80-208",
    doi = "10.1143/PTP.64.1103",
    journal = "Prog. Theor. Phys.",
    volume = "64",
    pages = "1103",
    year = "1980"
}

@article{Abdullahi:2022jlv,
    author = "Abdullahi, Asli M. and others",
    title = "{The present and future status of heavy neutral leptons}",
    eprint = "2203.08039",
    archivePrefix = "arXiv",
    primaryClass = "hep-ph",
    reportNumber = "FERMILAB-CONF-22-184-T-V",
    doi = "10.1088/1361-6471/ac98f9",
    journal = "J. Phys. G",
    volume = "50",
    number = "2",
    pages = "020501",
    year = "2023"
}

@article{LSND:2001aii,
    author = "Aguilar, A. and others",
    collaboration = "LSND",
    title = "{Evidence for neutrino oscillations from the observation of $\bar{\nu}_e$ appearance in a $\bar{\nu}_\mu$
 beam}",
    eprint = "hep-ex/0104049",
    archivePrefix = "arXiv",
    doi = "10.1103/PhysRevD.64.112007",
    journal = "Phys. Rev. D",
    volume = "64",
    pages = "112007",
    year = "2001"
}

@article{MiniBooNE:2007uho,
    author = "Aguilar-Arevalo, A. A. and others",
    collaboration = "MiniBooNE",
    title = "{A Search for Electron Neutrino Appearance at the $\Delta m^2 \sim 1 eV^2$ Scale}",
    eprint = "0704.1500",
    archivePrefix = "arXiv",
    primaryClass = "hep-ex",
    reportNumber = "FERMILAB-PUB-07-085-E, LA-UR-07-2246",
    doi = "10.1103/PhysRevLett.98.231801",
    journal = "Phys. Rev. Lett.",
    volume = "98",
    pages = "231801",
    year = "2007"
}

@article{Mention:2011rk,
    author = "Mention, G. and Fechner, M. and Lasserre, Th. and Mueller, Th. A. and Lhuillier, D. and Cribier, M. and Letourneau, A.",
    title = "{The Reactor Antineutrino Anomaly}",
    eprint = "1101.2755",
    archivePrefix = "arXiv",
    primaryClass = "hep-ex",
    doi = "10.1103/PhysRevD.83.073006",
    journal = "Phys. Rev. D",
    volume = "83",
    pages = "073006",
    year = "2011"
}

@article{Abdurashitov:1996dp,
    author = "Abdurashitov, Dzh. N. and others",
    title = "{The Russian-American gallium experiment (SAGE) Cr neutrino source measurement}",
    doi = "10.1103/PhysRevLett.77.4708",
    journal = "Phys. Rev. Lett.",
    volume = "77",
    pages = "4708--4711",
    year = "1996"
}

@article{Antusch:2016ejd,
    author = "Antusch, Stefan and Cazzato, Eros and Fischer, Oliver",
    title = "{Sterile neutrino searches at future $e^-e^+$, $pp$, and $e^-p$ colliders}",
    eprint = "1612.02728",
    archivePrefix = "arXiv",
    primaryClass = "hep-ph",
    doi = "10.1142/S0217751X17500786",
    journal = "Int. J. Mod. Phys. A",
    volume = "32",
    number = "14",
    pages = "1750078",
    year = "2017"
}

@article{ATLAS:2019kpx,
    author = "Aad, Georges and others",
    collaboration = "ATLAS",
    title = "{Search for heavy neutral leptons in decays of $W$ bosons produced in 13 TeV $pp$ collisions using prompt and displaced signatures with the ATLAS detector}",
    eprint = "1905.09787",
    archivePrefix = "arXiv",
    primaryClass = "hep-ex",
    reportNumber = "CERN-EP-2019-071",
    doi = "10.1007/JHEP10(2019)265",
    journal = "JHEP",
    volume = "10",
    pages = "265",
    year = "2019"
}

@article{Yang:2023ice,
    author = "Yang, Hao and Long, Bingwei and Qiao, Cong-Feng",
    title = "{Hunting for sterile neutrino with future collider signatures}",
    eprint = "2309.16233",
    archivePrefix = "arXiv",
    primaryClass = "hep-ph",
    doi = "10.1016/j.nuclphysb.2024.116576",
    journal = "Nucl. Phys. B",
    volume = "1004",
    pages = "116576",
    year = "2024"
}

@article{DeVries:2020jbs,
    author = {De Vries, Jordy and Dreiner, Herbert K. and G{\"u}nther, Julian Y. and Wang, Zeren Simon and Zhou, Guanghui},
    title = "{Long-lived Sterile Neutrinos at the LHC in Effective Field Theory}",
    eprint = "2010.07305",
    archivePrefix = "arXiv",
    primaryClass = "hep-ph",
    reportNumber = "APCTP Pre2020-027, BONN-TH-2020-10, RBRC-1328",
    doi = "10.1007/JHEP03(2021)148",
    journal = "JHEP",
    volume = "03",
    pages = "148",
    year = "2021"
}

@article{Avignone:2007fu,
    author = "Avignone, III, Frank T. and Elliott, Steven R. and Engel, Jonathan",
    title = "{Double Beta Decay, Majorana Neutrinos, and Neutrino Mass}",
    eprint = "0708.1033",
    archivePrefix = "arXiv",
    primaryClass = "nucl-ex",
    reportNumber = "LA-UR-07-3577",
    doi = "10.1103/RevModPhys.80.481",
    journal = "Rev. Mod. Phys.",
    volume = "80",
    pages = "481--516",
    year = "2008"
}

@article{Giuliani:2012zu,
    author = "Giuliani, Andrea and Poves, Alfredo",
    title = "{Neutrinoless Double-Beta Decay}",
    doi = "10.1155/2012/857016",
    journal = "Adv. High Energy Phys.",
    volume = "2012",
    pages = "857016",
    year = "2012"
}

@article{Harz:2021psp,
    author = "Harz, Julia and Ramsey-Musolf, Michael J. and Shen, Tianyang and Quiroga, Sebasti{\'a}n Urrutia",
    title = "{TeV-scale lepton number violation: Connecting leptogenesis, neutrinoless double beta decay, and colliders}",
    eprint = "2106.10838",
    archivePrefix = "arXiv",
    primaryClass = "hep-ph",
    reportNumber = "ACFI-T21-08, TUM-HEP-1346/21",
    doi = "10.1103/PhysRevD.110.035024",
    journal = "Phys. Rev. D",
    volume = "110",
    number = "3",
    pages = "035024",
    year = "2024"
}

@article{deVries:2024rfh,
    author = "de Vries, J. and Drewes, M. and Georis, Y. and Klari{\'c}, J. and Plakkot, V.",
    title = "{Confronting the low-scale seesaw and leptogenesis with neutrinoless double beta decay}",
    eprint = "2407.10560",
    archivePrefix = "arXiv",
    primaryClass = "hep-ph",
    reportNumber = "IRMP-CP3-24-20, ZTF-EP-24-10",
    doi = "10.1007/JHEP05(2025)090",
    journal = "JHEP",
    volume = "05",
    pages = "090",
    year = "2025"
}

@article{Dolinski:2019nrj,
    author = "Dolinski, Michelle J. and Poon, Alan W. P. and Rodejohann, Werner",
    title = "{Neutrinoless Double-Beta Decay: Status and Prospects}",
    eprint = "1902.04097",
    archivePrefix = "arXiv",
    primaryClass = "nucl-ex",
    doi = "10.1146/annurev-nucl-101918-023407",
    journal = "Ann. Rev. Nucl. Part. Sci.",
    volume = "69",
    pages = "219--251",
    year = "2019"
}

@article{DellOro:2016tmg,
    author = "Dell'Oro, Stefano and Marcocci, Simone and Viel, Matteo and Vissani, Francesco",
    title = "{Neutrinoless double beta decay: 2015 review}",
    eprint = "1601.07512",
    archivePrefix = "arXiv",
    primaryClass = "hep-ph",
    doi = "10.1155/2016/2162659",
    journal = "Adv. High Energy Phys.",
    volume = "2016",
    pages = "2162659",
    year = "2016"
}

@article{nEXO:2021ujk,
    author = "Adhikari, G. and others",
    collaboration = "nEXO",
    title = "{nEXO: neutrinoless double beta decay search beyond 10$^{28}$ year half-life sensitivity}",
    eprint = "2106.16243",
    archivePrefix = "arXiv",
    primaryClass = "nucl-ex",
    doi = "10.1088/1361-6471/ac3631",
    journal = "J. Phys. G",
    volume = "49",
    number = "1",
    pages = "015104",
    year = "2022"
}

@article{LEGEND:2021bnm,
    author = "Abgrall, N. and others",
    collaboration = "LEGEND",
    title = "{The Large Enriched Germanium Experiment for Neutrinoless $\beta\beta$ Decay}: {LEGEND-1000 Preconceptual Design Report}",
    eprint = "2107.11462",
    archivePrefix = "arXiv",
    primaryClass = "physics.ins-det",
    month = "7",
    year = "2021"
}

@article{CUPID:2022wpt,
    author = "Armatol, A. and others",
    collaboration = "CUPID",
    title = "{Toward CUPID-1T}",
    eprint = "2203.08386",
    archivePrefix = "arXiv",
    primaryClass = "nucl-ex",
    month = "3",
    year = "2022"
}

@article{Blennow:2010th,
    author = "Blennow, Mattias and Fernandez-Martinez, Enrique and Lopez-Pavon, Jacobo and Menendez, Javier",
    title = "{Neutrinoless double beta decay in seesaw models}",
    eprint = "1005.3240",
    archivePrefix = "arXiv",
    primaryClass = "hep-ph",
    reportNumber = "MPP-2010-40, IFT-UAM-CSIC-10-26, FTUAM-10-06, EURONU-WP6-10-18",
    doi = "10.1007/JHEP07(2010)096",
    journal = "JHEP",
    volume = "07",
    pages = "096",
    year = "2010"
}

@article{Mitra:2011qr,
    author = "Mitra, Manimala and Senjanovic, Goran and Vissani, Francesco",
    title = "{Neutrinoless Double Beta Decay and Heavy Sterile Neutrinos}",
    eprint = "1108.0004",
    archivePrefix = "arXiv",
    primaryClass = "hep-ph",
    doi = "10.1016/j.nuclphysb.2011.10.035",
    journal = "Nucl. Phys. B",
    volume = "856",
    pages = "26--73",
    year = "2012"
}

@article{Asaka:2016zib,
    author = "Asaka, Takehiko and Eijima, Shintaro and Ishida, Hiroyuki",
    title = "{On neutrinoless double beta decay in the $\nu$MSM}",
    eprint = "1606.06686",
    archivePrefix = "arXiv",
    primaryClass = "hep-ph",
    reportNumber = "SU-HET-05-2016",
    doi = "10.1016/j.physletb.2016.09.044",
    journal = "Phys. Lett. B",
    volume = "762",
    pages = "371--375",
    year = "2016"
}

@article{Faessler:2014kka,
    author = "Faessler, Amand and Gonz{\'a}lez, Marcela and Kovalenko, Sergey and {\v{S}}imkovic, Fedor",
    title = "{Arbitrary mass Majorana neutrinos in neutrinoless double beta decay}",
    eprint = "1408.6077",
    archivePrefix = "arXiv",
    primaryClass = "hep-ph",
    doi = "10.1103/PhysRevD.90.096010",
    journal = "Phys. Rev. D",
    volume = "90",
    number = "9",
    pages = "096010",
    year = "2014"
}

@article{Dekens:2024hlz,
    author = "Dekens, W. and de Vries, J. and Castillo, D. and Men{\'e}ndez, J. and Mereghetti, E. and Plakkot, V. and Soriano, P. and Zhou, G.",
    title = "{Neutrinoless double beta decay rates in the presence of light sterile neutrinos}",
    eprint = "2402.07993",
    archivePrefix = "arXiv",
    primaryClass = "hep-ph",
    reportNumber = "LA-UR-24-21117, INT-PUB-24-007",
    doi = "10.1007/JHEP09(2024)201",
    journal = "JHEP",
    volume = "09",
    pages = "201",
    year = "2024"
}

@article{Dekens:2023iyc,
    author = "Dekens, Wouter and de Vries, Jordy and Mereghetti, Emanuele and Men{\'e}ndez, Javier and Soriano, Pablo and Zhou, Guanghui",
    title = "{Neutrinoless double-{\ensuremath{\beta}} decay in the neutrino-extended standard model}",
    eprint = "2303.04168",
    archivePrefix = "arXiv",
    primaryClass = "hep-ph",
    reportNumber = "LA-UR-23-22287, INT-PUB-23-009",
    doi = "10.1103/PhysRevC.108.045501",
    journal = "Phys. Rev. C",
    volume = "108",
    number = "4",
    pages = "045501",
    year = "2023"
}

@article{Huang:2019qvq,
    author = "Huang, Guo-Yuan and Zhou, Shun",
    title = "{Impact of an eV-mass sterile neutrino on the neutrinoless double-beta decays: A Bayesian analysis}",
    eprint = "1902.03839",
    archivePrefix = "arXiv",
    primaryClass = "hep-ph",
    doi = "10.1016/j.nuclphysb.2019.114691",
    journal = "Nucl. Phys. B",
    volume = "945",
    pages = "114691",
    year = "2019"
}

@article{Caurier:2007qn,
    author = "Caurier, E. and Nowacki, F. and Poves, A.",
    title = "{Nuclear Structure Aspects of the Neutrinoless Double Beta Decay}",
    eprint = "0709.0277",
    archivePrefix = "arXiv",
    primaryClass = "nucl-th",
    doi = "10.1140/epja/i2007-10527-x",
    journal = "Eur. Phys. J. A",
    volume = "36",
    pages = "195--200",
    year = "2008"
}

@article{Brown:2001zz,
    author = "Brown, B. A.",
    title = "{The nuclear shell model towards the drip lines}",
    doi = "10.1016/S0146-6410(01)00159-4",
    journal = "Prog. Part. Nucl. Phys.",
    volume = "47",
    pages = "517--599",
    year = "2001"
}

@article{Pantis:1996py,
    author = "Pantis, G. and Simkovic, F. and Vergados, J. D. and Faessler, Amand",
    title = "{Neutrinoless double beta decay within QRPA with proton - neutron pairing}",
    eprint = "nucl-th/9612036",
    archivePrefix = "arXiv",
    doi = "10.1103/PhysRevC.53.695",
    journal = "Phys. Rev. C",
    volume = "53",
    pages = "695--707",
    year = "1996"
}

@article{Bobyk:2000dw,
    author = "Bobyk, A. and Kaminski, W. A. and Simkovic, F.",
    title = "{Neutrinoless double beta decay within selfconsistent renormalized quasiparticle random phase approximation and inclusion of induced nucleon currents}",
    eprint = "nucl-th/0012010",
    archivePrefix = "arXiv",
    doi = "10.1103/PhysRevC.63.051301",
    journal = "Phys. Rev. C",
    volume = "63",
    pages = "051301",
    year = "2001"
}

@article{Arima:1981hp,
    author = "Arima, A. and Iachello, F.",
    title = "{The interacting boson model}",
    doi = "10.1146/annurev.ns.31.120181.000451",
    journal = "Ann. Rev. Nucl. Part. Sci.",
    volume = "31",
    pages = "75--105",
    year = "1981"
}

@article{Barea:2009zza,
    author = "Barea, J. and Iachello, F.",
    title = "{Neutrinoless double-beta decay in the microscopic interacting boson model}",
    doi = "10.1103/PhysRevC.79.044301",
    journal = "Phys. Rev. C",
    volume = "79",
    pages = "044301",
    year = "2009"
}

@article{Rodriguez:2010mn,
    author = "Rodriguez, Tomas R. and Martinez-Pinedo, G.",
    title = "{Energy density functional study of nuclear matrix elements for neutrinoless $\beta\beta$ decay}",
    eprint = "1008.5260",
    archivePrefix = "arXiv",
    primaryClass = "nucl-th",
    doi = "10.1103/PhysRevLett.105.252503",
    journal = "Phys. Rev. Lett.",
    volume = "105",
    pages = "252503",
    year = "2010"
}

@article{LopezVaquero:2013yji,
    author = "L{\'o}pez Vaquero, Nuria and Rodr{\'\i}guez, Tom{\'a}s R. and Egido, J. Luis",
    title = "{Shape and pairing fluctuations effects on neutrinoless double beta decay nuclear matrix elements}",
    eprint = "1401.0650",
    archivePrefix = "arXiv",
    primaryClass = "nucl-th",
    doi = "10.1103/PhysRevLett.111.142501",
    journal = "Phys. Rev. Lett.",
    volume = "111",
    number = "14",
    pages = "142501",
    year = "2013"
}

@article{Weinberg:1990rz,
    author = "Weinberg, Steven",
    title = "{Nuclear forces from chiral Lagrangians}",
    reportNumber = "UTTG-31-90",
    doi = "10.1016/0370-2693(90)90938-3",
    journal = "Phys. Lett. B",
    volume = "251",
    pages = "288--292",
    year = "1990"
}

@article{Weinberg:1991um,
    author = "Weinberg, Steven",
    title = "{Effective chiral Lagrangians for nucleon - pion interactions and nuclear forces}",
    reportNumber = "UTTG-03-91",
    doi = "10.1016/0550-3213(91)90231-L",
    journal = "Nucl. Phys. B",
    volume = "363",
    pages = "3--18",
    year = "1991"
}

@article{Ordonez:1992xp,
    author = "Ordonez, C. and van Kolck, U.",
    title = "{Chiral lagrangians and nuclear forces}",
    reportNumber = "UTTG-01-92",
    doi = "10.1016/0370-2693(92)91404-W",
    journal = "Phys. Lett. B",
    volume = "291",
    pages = "459--464",
    year = "1992"
}

@article{Cirigliano:2017tvr,
    author = "Cirigliano, Vincenzo and Dekens, Wouter and Mereghetti, Emanuele and Walker-Loud, Andr{\'e}",
    title = "{Neutrinoless double-{\ensuremath{\beta}} decay in effective field theory: The light-Majorana neutrino-exchange mechanism}",
    eprint = "1710.01729",
    archivePrefix = "arXiv",
    primaryClass = "hep-ph",
    reportNumber = "LA-UR-17-28401",
    doi = "10.1103/PhysRevC.97.065501",
    journal = "Phys. Rev. C",
    volume = "97",
    number = "6",
    pages = "065501",
    year = "2018",
    note = "[Erratum: Phys.Rev.C 100, 019903 (2019)]"
}

@article{Brase:2021uny,
    author = "Brase, Catharina and Men{\'e}ndez, Javier and Coello P{\'e}rez, Eduardo Antonio and Schwenk, Achim",
    title = "{Neutrinoless double-{\ensuremath{\beta}} decay from an effective field theory for heavy nuclei}",
    eprint = "2108.11805",
    archivePrefix = "arXiv",
    primaryClass = "nucl-th",
    doi = "10.1103/PhysRevC.106.034309",
    journal = "Phys. Rev. C",
    volume = "106",
    number = "3",
    pages = "034309",
    year = "2022"
}

@article{Das:2023aic,
    author = "Das, Debashree Priyadarsini and Mishra, Sasmita",
    title = "{Study of neutrinoless double beta decay in the Standard Model extended with sterile neutrinos}",
    eprint = "2310.13353",
    archivePrefix = "arXiv",
    primaryClass = "hep-ph",
    doi = "10.1140/epjc/s10052-024-13024-w",
    journal = "Eur. Phys. J. C",
    volume = "84",
    number = "7",
    pages = "683",
    year = "2024"
}

@article{Kotila:2012zza,
    author = "Kotila, J. and Iachello, F.",
    title = "{Phase space factors for double-$\beta$ decay}",
    eprint = "1209.5722",
    archivePrefix = "arXiv",
    primaryClass = "nucl-th",
    doi = "10.1103/PhysRevC.85.034316",
    journal = "Phys. Rev. C",
    volume = "85",
    pages = "034316",
    year = "2012"
}

@article{ParticleDataGroup:2024cfk,
    author = "Navas, S. and others",
    collaboration = "Particle Data Group",
    title = "{Review of particle physics}",
    doi = "10.1103/PhysRevD.110.030001",
    journal = "Phys. Rev. D",
    volume = "110",
    number = "3",
    pages = "030001",
    year = "2024"
}

@article{Beneke:1997zp,
    author = "Beneke, M. and Smirnov, Vladimir A.",
    title = "{Asymptotic expansion of Feynman integrals near threshold}",
    eprint = "hep-ph/9711391",
    archivePrefix = "arXiv",
    reportNumber = "CERN-TH-97-315",
    doi = "10.1016/S0550-3213(98)00138-2",
    journal = "Nucl. Phys. B",
    volume = "522",
    pages = "321--344",
    year = "1998"
}

@article{Machleidt:2011zz,
    author = "Machleidt, R. and Entem, D. R.",
    title = "{Chiral effective field theory and nuclear forces}",
    eprint = "1105.2919",
    archivePrefix = "arXiv",
    primaryClass = "nucl-th",
    doi = "10.1016/j.physrep.2011.02.001",
    journal = "Phys. Rept.",
    volume = "503",
    pages = "1--75",
    year = "2011"
}

@article{Cirigliano:2019vdj,
    author = "Cirigliano, V. and Dekens, W. and De Vries, J. and Graesser, M. L. and Mereghetti, E. and Pastore, S. and Piarulli, M. and Van Kolck, U. and Wiringa, R. B.",
    title = "{Renormalized approach to neutrinoless double- $\beta$ decay}",
    eprint = "1907.11254",
    archivePrefix = "arXiv",
    primaryClass = "nucl-th",
    doi = "10.1103/PhysRevC.100.055504",
    journal = "Phys. Rev. C",
    volume = "100",
    number = "5",
    pages = "055504",
    year = "2019"
}

@article{Cirigliano:2018hja,
    author = "Cirigliano, Vincenzo and Dekens, Wouter and De Vries, Jordy and Graesser, Michael L. and Mereghetti, Emanuele and Pastore, Saori and Van Kolck, Ubirajara",
    title = "{New Leading Contribution to Neutrinoless Double-{\ensuremath{\beta}} Decay}",
    eprint = "1802.10097",
    archivePrefix = "arXiv",
    primaryClass = "hep-ph",
    reportNumber = "LA-UR-18-21404, NIKHEF-2018-010",
    doi = "10.1103/PhysRevLett.120.202001",
    journal = "Phys. Rev. Lett.",
    volume = "120",
    number = "20",
    pages = "202001",
    year = "2018"
}

@article{Nicholson:2018mwc,
    author = "Nicholson, A. and others",
    title = "{Heavy physics contributions to neutrinoless double beta decay from QCD}",
    eprint = "1805.02634",
    archivePrefix = "arXiv",
    primaryClass = "nucl-th",
    reportNumber = "LLNL-JRNL-751220, RBRC-1266, RIKEN-ITHEMS-REPORT-18, RIKEN-iTHEMS-Report-18, BNL-209118-2018-JAAM",
    doi = "10.1103/PhysRevLett.121.172501",
    journal = "Phys. Rev. Lett.",
    volume = "121",
    number = "17",
    pages = "172501",
    year = "2018"
}

@article{Jokiniemi:2021qqv,
    author = "Jokiniemi, Lotta and Soriano, Pablo and Men{\'e}ndez, Javier",
    title = "{Impact of the leading-order short-range nuclear matrix element on the neutrinoless double-beta decay of medium-mass and heavy nuclei}",
    eprint = "2107.13354",
    archivePrefix = "arXiv",
    primaryClass = "nucl-th",
    doi = "10.1016/j.physletb.2021.136720",
    journal = "Phys. Lett. B",
    volume = "823",
    pages = "136720",
    year = "2021"
}

@article{Cirigliano:2018yza,
    author = "Cirigliano, V. and Dekens, W. and de Vries, J. and Graesser, M. L. and Mereghetti, E.",
    title = "{A neutrinoless double beta decay master formula from effective field theory}",
    eprint = "1806.02780",
    archivePrefix = "arXiv",
    primaryClass = "hep-ph",
    reportNumber = "LA-UR-18-24895, Nikhef 2018-023, NIKHEF-2018-023, DESY-18-072",
    doi = "10.1007/JHEP12(2018)097",
    journal = "JHEP",
    volume = "12",
    pages = "097",
    year = "2018"
}

@article{Xing:2007zj,
    author = "Xing, Zhi-zhong",
    title = "{Correlation between the Charged Current Interactions of Light and Heavy Majorana Neutrinos}",
    eprint = "0709.2220",
    archivePrefix = "arXiv",
    primaryClass = "hep-ph",
    doi = "10.1016/j.physletb.2008.01.038",
    journal = "Phys. Lett. B",
    volume = "660",
    pages = "515--521",
    year = "2008"
}

@article{Xing:2011ur,
    author = "Xing, Zhi-zhong",
    title = "{A full parametrization of the 6 X 6 flavor mixing matrix in the presence of three light or heavy sterile neutrinos}",
    eprint = "1110.0083",
    archivePrefix = "arXiv",
    primaryClass = "hep-ph",
    doi = "10.1103/PhysRevD.85.013008",
    journal = "Phys. Rev. D",
    volume = "85",
    pages = "013008",
    year = "2012"
}

@article{Esteban:2024eli,
    author = "Esteban, Ivan and Gonzalez-Garcia, M. C. and Maltoni, Michele and Martinez-Soler, Ivan and Pinheiro, Jo{\~a}o Paulo and Schwetz, Thomas",
    title = "{NuFit-6.0: updated global analysis of three-flavor neutrino oscillations}",
    eprint = "2410.05380",
    archivePrefix = "arXiv",
    primaryClass = "hep-ph",
    reportNumber = "IFT-UAM/CSIC-24-140, YITP-SB-2024-24, IPPP/24/64, IPPP/24/64, IFT-UAM/CSIC-24-140, YITP-SB-2024-24",
    doi = "10.1007/JHEP12(2024)216",
    journal = "JHEP",
    volume = "12",
    pages = "216",
    year = "2024"
}

@article{Bolton:2019pcu,
    author = "Bolton, Patrick D. and Deppisch, Frank F. and Bhupal Dev, P. S.",
    title = "{Neutrinoless double beta decay versus other probes of heavy sterile neutrinos}",
    eprint = "1912.03058",
    archivePrefix = "arXiv",
    primaryClass = "hep-ph",
    doi = "10.1007/JHEP03(2020)170",
    journal = "JHEP",
    volume = "03",
    pages = "170",
    year = "2020"
}

@article{Xing:2009in,
    author = "Xing, Zhi-zhong",
    editor = "Kobayashi, Tatsuo and Kugo, Taichiro",
    title = "{Naturalness and Testability of TeV Seesaw Mechanisms}",
    eprint = "0905.3903",
    archivePrefix = "arXiv",
    primaryClass = "hep-ph",
    doi = "10.1143/PTPS.180.112",
    journal = "Prog. Theor. Phys. Suppl.",
    volume = "180",
    pages = "112--127",
    year = "2009"
}

@article{Atre:2009rg,
    author = "Atre, Anupama and Han, Tao and Pascoli, Silvia and Zhang, Bin",
    title = "{The Search for Heavy Majorana Neutrinos}",
    eprint = "0901.3589",
    archivePrefix = "arXiv",
    primaryClass = "hep-ph",
    reportNumber = "FERMILAB-PUB-08-086-T, NSF-KITP-08-54, MADPH-06-1466, DCPT-07-198, IPPP-07-99",
    doi = "10.1088/1126-6708/2009/05/030",
    journal = "JHEP",
    volume = "05",
    pages = "030",
    year = "2009"
}

@article{Cabrera:2023rcy,
    author = "Cabrera, E. and Cogollo, D. and Pires, C. A. de S.",
    title = "{Naturally low scale type I seesaw mechanism and its viability in the 3-3-1 model with right-handed neutrinos}",
    eprint = "2304.14443",
    archivePrefix = "arXiv",
    primaryClass = "hep-ph",
    doi = "10.1016/j.nuclphysb.2023.116372",
    journal = "Nucl. Phys. B",
    volume = "996",
    pages = "116372",
    year = "2023"
}

@article{LEGEND:2017cdu,
    author = "Abgrall, N. and others",
    editor = "Civitarese, Osvaldo and Stekl, Ivan and Suhonen, Jouni",
    collaboration = "LEGEND",
    title = "{The Large Enriched Germanium Experiment for Neutrinoless Double Beta Decay (LEGEND)}",
    eprint = "1709.01980",
    archivePrefix = "arXiv",
    primaryClass = "physics.ins-det",
    doi = "10.1063/1.5007652",
    journal = "AIP Conf. Proc.",
    volume = "1894",
    number = "1",
    pages = "020027",
    year = "2017"
}

@article{nEXO:2017nam,
    author = "Albert, J. B. and others",
    collaboration = "nEXO",
    title = "{Sensitivity and Discovery Potential of nEXO to Neutrinoless Double Beta Decay}",
    eprint = "1710.05075",
    archivePrefix = "arXiv",
    primaryClass = "nucl-ex",
    reportNumber = "LLNL-JRNL-737682",
    doi = "10.1103/PhysRevC.97.065503",
    journal = "Phys. Rev. C",
    volume = "97",
    number = "6",
    pages = "065503",
    year = "2018"
}

@article{nEXO:2018ylp,
    author = "Kharusi, S. Al and others",
    collaboration = "nEXO",
    title = "{nEXO Pre-Conceptual Design Report}",
    eprint = "1805.11142",
    archivePrefix = "arXiv",
    primaryClass = "physics.ins-det",
    month = "5",
    year = "2018"
}

@article{CUPID:2019imh,
    author = "Armstrong, W. R. and others",
    collaboration = "CUPID",
    title = "{CUPID pre-CDR}",
    eprint = "1907.09376",
    archivePrefix = "arXiv",
    primaryClass = "physics.ins-det",
    month = "7",
    year = "2019"
}

@article{Armengaud:2019loe,
    author = "Armengaud, E. and others",
    title = "{The CUPID-Mo experiment for neutrinoless double-beta decay: performance and prospects}",
    eprint = "1909.02994",
    archivePrefix = "arXiv",
    primaryClass = "physics.ins-det",
    doi = "10.1140/epjc/s10052-019-7578-6",
    journal = "Eur. Phys. J. C",
    volume = "80",
    number = "1",
    pages = "44",
    year = "2020"
}

@article{Merle:2013ibc,
    author = "Merle, Alexander and Niro, Viviana",
    title = "{Influence of a keV sterile neutrino on neutrinoless double beta decay: How things changed in recent years}",
    eprint = "1302.2032",
    archivePrefix = "arXiv",
    primaryClass = "hep-ph",
    reportNumber = "UB-ECM-PF-87-13, ICCUB-13-042",
    doi = "10.1103/PhysRevD.88.113004",
    journal = "Phys. Rev. D",
    volume = "88",
    number = "11",
    pages = "113004",
    year = "2013"
}

@article{Asaka:2005an,
    author = "Asaka, Takehiko and Blanchet, Steve and Shaposhnikov, Mikhail",
    title = "{The nuMSM, dark matter and neutrino masses}",
    eprint = "hep-ph/0503065",
    archivePrefix = "arXiv",
    doi = "10.1016/j.physletb.2005.09.070",
    journal = "Phys. Lett. B",
    volume = "631",
    pages = "151--156",
    year = "2005"
}

@article{Merle:2015vzu,
    author = "Merle, Alexander and Schneider, Aurel and Totzauer, Maximilian",
    title = "{Dodelson-Widrow Production of Sterile Neutrino Dark Matter with Non-Trivial Initial Abundance}",
    eprint = "1512.05369",
    archivePrefix = "arXiv",
    primaryClass = "hep-ph",
    reportNumber = "MPP-2015-302",
    doi = "10.1088/1475-7516/2016/04/003",
    journal = "JCAP",
    volume = "04",
    pages = "003",
    year = "2016"
}

@article{Shi:1998km,
    author = "Shi, Xiang-Dong and Fuller, George M.",
    title = "{A New dark matter candidate: Nonthermal sterile neutrinos}",
    eprint = "astro-ph/9810076",
    archivePrefix = "arXiv",
    doi = "10.1103/PhysRevLett.82.2832",
    journal = "Phys. Rev. Lett.",
    volume = "82",
    pages = "2832--2835",
    year = "1999"
}

@article{Lindner:2010wr,
    author = "Lindner, Manfred and Merle, Alexander and Niro, Viviana",
    title = "{Soft $L_e - L_\mu - L_\tau$ flavour symmetry breaking and sterile neutrino keV Dark Matter}",
    eprint = "1011.4950",
    archivePrefix = "arXiv",
    primaryClass = "hep-ph",
    doi = "10.1088/1475-7516/2011/01/034",
    journal = "JCAP",
    volume = "01",
    pages = "034",
    year = "2011",
    note = "[Erratum: JCAP 07, E01 (2014)]"
}

@article{CarcamoHernandez:2019kjy,
    author = "C{\'a}rcamo Hern{\'a}ndez, A. E. and Gonz{\'a}lez, Marcela and Neill, Nicol{\'a}s A.",
    title = "{Low scale type I seesaw model for lepton masses and mixings}",
    eprint = "1906.00978",
    archivePrefix = "arXiv",
    primaryClass = "hep-ph",
    doi = "10.1103/PhysRevD.101.035005",
    journal = "Phys. Rev. D",
    volume = "101",
    number = "3",
    pages = "035005",
    year = "2020"
}

@article{Barry:2011wb,
    author = "Barry, James and Rodejohann, Werner and Zhang, He",
    title = "{Light Sterile Neutrinos: Models and Phenomenology}",
    eprint = "1105.3911",
    archivePrefix = "arXiv",
    primaryClass = "hep-ph",
    doi = "10.1007/JHEP07(2011)091",
    journal = "JHEP",
    volume = "07",
    pages = "091",
    year = "2011"
}

@article{Conrad:2012qt,
    author = "Conrad, J. M. and Ignarra, C. M. and Karagiorgi, G. and Shaevitz, M. H. and Spitz, J.",
    title = "{Sterile Neutrino Fits to Short Baseline Neutrino Oscillation Measurements}",
    eprint = "1207.4765",
    archivePrefix = "arXiv",
    primaryClass = "hep-ex",
    doi = "10.1155/2013/163897",
    journal = "Adv. High Energy Phys.",
    volume = "2013",
    pages = "163897",
    year = "2013"
}

@article{Danilov:2022str,
    author = "Danilov, Mikhail",
    title = "{Review of sterile neutrino searches at very short-baseline reactor experiments}",
    eprint = "2203.03042",
    archivePrefix = "arXiv",
    primaryClass = "hep-ex",
    doi = "10.1088/1402-4896/ac81fd",
    journal = "Phys. Scripta",
    volume = "97",
    number = "9",
    pages = "094001",
    year = "2022"
}

@article{Katori:2014vka,
    author = "Katori, Teppei",
    title = "{Short Baseline Neutrino Oscillation Experiments}",
    eprint = "1404.6882",
    archivePrefix = "arXiv",
    primaryClass = "hep-ph",
    doi = "10.1088/1742-6596/598/1/012006",
    journal = "J. Phys. Conf. Ser.",
    volume = "598",
    number = "1",
    pages = "012006",
    year = "2015"
}

@article{Boyarsky:2009ix,
    author = "Boyarsky, Alexey and Ruchayskiy, Oleg and Shaposhnikov, Mikhail",
    title = "{The Role of sterile neutrinos in cosmology and astrophysics}",
    eprint = "0901.0011",
    archivePrefix = "arXiv",
    primaryClass = "hep-ph",
    doi = "10.1146/annurev.nucl.010909.083654",
    journal = "Ann. Rev. Nucl. Part. Sci.",
    volume = "59",
    pages = "191--214",
    year = "2009"
}

@article{Antusch:2015mia,
    author = "Antusch, Stefan and Fischer, Oliver",
    title = "{Testing sterile neutrino extensions of the Standard Model at future lepton colliders}",
    eprint = "1502.05915",
    archivePrefix = "arXiv",
    primaryClass = "hep-ph",
    reportNumber = "MPP-2015-24",
    doi = "10.1007/JHEP05(2015)053",
    journal = "JHEP",
    volume = "05",
    pages = "053",
    year = "2015"
}

@article{Abada:2022wvh,
    author = "Abada, Asmaa and Escribano, Pablo and Marcano, Xabier and Piazza, Gioacchino",
    title = "{Collider searches for heavy neutral leptons: beyond simplified scenarios}",
    eprint = "2208.13882",
    archivePrefix = "arXiv",
    primaryClass = "hep-ph",
    reportNumber = "IFT-UAM/CSIC-22-98",
    doi = "10.1140/epjc/s10052-022-11011-7",
    journal = "Eur. Phys. J. C",
    volume = "82",
    number = "11",
    pages = "1030",
    year = "2022"
}
\end{document}